\documentclass[twocolumn,amsmath,amssymb,prd]{revtex4}

\def\al{\alpha}
\def\be{\beta}
\def\ga{\gamma}
\def\de{\delta}

\def\th{\theta}

\def\si{\sigma}

\def\ch{\chi}
\def\ps{\psi}
\def\om{\omega}

\def\De{\Delta}

\def\cH{{\cal H}}
\def\cl{{\cal L}}

\def\fr#1#2{{{#1}\over{#2}}}
\def\frac#1#2{{\textstyle{{#1}\over{#2}}}}
\def\half{{\textstyle{1\over 2}}}
\def\ol{\overline}
\def\prt{\partial}

\def\lsim{\mathrel{\rlap{\lower4pt\hbox{\hskip1pt$\sim$}}
    \raise1pt\hbox{$<$}}}
\def\gsim{\mathrel{\rlap{\lower4pt\hbox{\hskip1pt$\sim$}}
    \raise1pt\hbox{$>$}}}

\def\etal{{\it et al.}}
\def\nn{\nonumber}

\def\vev#1{\langle {#1}\rangle}

\newcommand{\beq}{\begin{equation}}
\newcommand{\eeq}{\end{equation}}
\newcommand{\bea}{\begin{eqnarray}}
\newcommand{\eea}{\end{eqnarray}}

\newcommand{\rf}[1]{(\ref{#1})}

\newcommand{\tab}[1]{Table~\ref{#1}}

\def\psb{\ol\ps{}}
\def\mbf#1{\boldsymbol #1}

\def\pvec{\mbf p}
\def\gavec{\mbf\ga}

\def\Q{\mathcal Q}
\def\S{\mathcal S}
\def\P{\mathcal P}
\def\V{\mathcal V}
\def\A{\mathcal A}
\def\T{\mathcal T}
\def\Qhat{\widehat\Q}
\def\Shat{\widehat\S}
\def\Phat{\widehat\P}
\def\Vhat{\widehat\V}
\def\Ahat{\widehat\A}
\def\That{\widehat\T}

\def\codt{\cos{\om_\oplus T_\oplus}}
\def\sodt{\sin{\om_\oplus T_\oplus}}
\def\ctodt{\cos{2\om_\oplus T_\oplus}}
\def\stodt{\sin{2\om_\oplus T_\oplus}}
\def\cthodt{\cos{3\om_\oplus T_\oplus}}
\def\sthodt{\sin{3\om_\oplus T_\oplus}}

\def\cmtemplate#1#2#3#4{{#1}^{#3}_{#4}}

\def\acm#1#2{\cmtemplate{a}{#1}{#2}{}}
\def\bcm#1#2{\cmtemplate{b}{#1}{#2}{}}
\def\ccm#1#2{\cmtemplate{c}{#1}{#2}{}}
\def\dcm#1#2{\cmtemplate{d}{#1}{#2}{}}
\def\ecm#1#2{\cmtemplate{e}{#1}{#2}{}}
\def\fcm#1#2{\cmtemplate{f}{#1}{#2}{}}
\def\gcm#1#2{\cmtemplate{g}{#1}{#2}{}}
\def\Hcm#1#2{\cmtemplate{H}{#1}{#2}{}}

\def\ctemplate#1#2#3#4{{#1}^{(#2)#3}_{#4}}
\def\mc#1#2{\ctemplate{m}{#1}{#2}{}}
\def\mfc#1#2{\ctemplate{m}{#1}{#2}{5}}
\def\ac#1#2{\ctemplate{a}{#1}{#2}{}}
\def\bc#1#2{\ctemplate{b}{#1}{#2}{}}
\def\cc#1#2{\ctemplate{c}{#1}{#2}{}}
\def\dc#1#2{\ctemplate{d}{#1}{#2}{}}
\def\ec#1#2{\ctemplate{e}{#1}{#2}{}}
\def\fc#1#2{\ctemplate{f}{#1}{#2}{}}
\def\gc#1#2{\ctemplate{g}{#1}{#2}{}}
\def\Hc#1#2{\ctemplate{H}{#1}{#2}{}}

\def\mcf#1#2{\ctemplate{m}{#1}{#2}{F}}
\def\mfcf#1#2{\ctemplate{m}{#1}{#2}{5F}}
\def\acf#1#2{\ctemplate{a}{#1}{#2}{F}}
\def\bcf#1#2{\ctemplate{b}{#1}{#2}{F}}
\def\ccf#1#2{\ctemplate{c}{#1}{#2}{F}}
\def\dcf#1#2{\ctemplate{d}{#1}{#2}{F}}
\def\ecf#1#2{\ctemplate{e}{#1}{#2}{F}}
\def\fcf#1#2{\ctemplate{f}{#1}{#2}{F}}
\def\gcf#1#2{\ctemplate{g}{#1}{#2}{F}}
\def\Hcf#1#2{\ctemplate{H}{#1}{#2}{F}}

\def\mcpf#1#2{\ctemplate{m}{#1}{#2}{\prt F}}
\def\mfcpf#1#2{\ctemplate{m}{#1}{#2}{5\prt F}}
\def\acpf#1#2{\ctemplate{a}{#1}{#2}{\prt F}}
\def\bcpf#1#2{\ctemplate{b}{#1}{#2}{\prt F}}

\def\Hcpf#1#2{\ctemplate{H}{#1}{#2}{\prt F}}

\def\cmtemplate#1#2#3#4{{#1}^{#3}_{#4}}

\def\acmw#1#2#3{\cmtemplate{a}{#1}{#2}{{#3}}}
\def\bcmw#1#2#3{\cmtemplate{b}{#1}{#2}{{#3}}}
\def\ccmw#1#2#3{\cmtemplate{c}{#1}{#2}{{#3}}}
\def\dcmw#1#2#3{\cmtemplate{d}{#1}{#2}{{#3}}}
\def\ecmw#1#2#3{\cmtemplate{e}{#1}{#2}{{#3}}}

\def\gcmw#1#2#3{\cmtemplate{g}{#1}{#2}{{#3}}}
\def\Hcmw#1#2#3{\cmtemplate{H}{#1}{#2}{{#3}}}

\def\ctemplate#1#2#3#4{{#1}^{(#2)#3}_{#4}}
\def\mcw#1#2#3{\ctemplate{m}{#1}{#2}{{#3}}}

\def\acw#1#2#3{\ctemplate{a}{#1}{#2}{{#3}}}
\def\bcw#1#2#3{\ctemplate{b}{#1}{#2}{{#3}}}
\def\ccw#1#2#3{\ctemplate{c}{#1}{#2}{{#3}}}
\def\dcw#1#2#3{\ctemplate{d}{#1}{#2}{{#3}}}
\def\ecw#1#2#3{\ctemplate{e}{#1}{#2}{{#3}}}

\def\gcw#1#2#3{\ctemplate{g}{#1}{#2}{{#3}}}
\def\Hcw#1#2#3{\ctemplate{H}{#1}{#2}{{#3}}}

\def\mcfw#1#2#3{\ctemplate{m}{#1}{#2}{F{,#3}}}

\def\acfw#1#2#3{\ctemplate{a}{#1}{#2}{F{,#3}}}
\def\bcfw#1#2#3{\ctemplate{b}{#1}{#2}{F{,#3}}}
\def\ccfw#1#2#3{\ctemplate{c}{#1}{#2}{F{,#3}}}
\def\dcfw#1#2#3{\ctemplate{d}{#1}{#2}{F{,#3}}}
\def\ecfw#1#2#3{\ctemplate{e}{#1}{#2}{F{,#3}}}

\def\gcfw#1#2#3{\ctemplate{g}{#1}{#2}{F{,#3}}}
\def\Hcfw#1#2#3{\ctemplate{H}{#1}{#2}{F{,#3}}}

\def\mn{{\mu\nu}}
\def\ma{{\mu\al}}
\def\mna{{\mu\nu\al}}
\def\ab{{\al\be}}
\def\bec{{\be\ga}}
\def\mab{{\mu\al\be}}
\def\mnab{{\mu\nu\al\be}}
\def\abc{{\al\be\ga}}

\def\mabc{{\mu\al\be\ga}}
\def\mnabc{{\mu\nu\al\be\ga}}

\def\m{m_\ps}
\def\mw{m_w}

\def\quar{\frac 1 4}

\def\cthodt{\cos{3\om_\oplus T_\oplus}}
\def\sthodt{\sin{3\om_\oplus T_\oplus}}

\def\ens{E_{n,\pm}}

\def\atw#1#2{{\widetilde a}_{#1}^{#2}}
\def\btw#1#2{{\widetilde b}_{#1}^{#2}}
\def\bftw#1#2{{\widetilde b}_{F,#1}^{#2}}
\def\mftw#1#2{{\widetilde m}_{F,#1}^{#2}}
\def\atws#1#2{{\widetilde a}_{#1}^{*#2}}
\def\btws#1#2{{\widetilde b}_{#1}^{*#2}}
\def\bftws#1#2{{\widetilde b}_{F,#1}^{*#2}}
\def\mftws#1#2{{\widetilde m}_{F,#1}^{*#2}}

\def\bptw#1#2{{\widetilde b}_{#1}^{\prime #2}}

\def\ctw#1#2{{\widetilde c}_{#1}^{#2}}
\def\bptws#1#2{{\widetilde b}_{#1}^{\prime *#2}}

\def\ctws#1#2{{\widetilde c}_{#1}^{*#2}}

\def\vos{\mathrel{\rlap{\lower0pt\hbox{\hskip0.5pt{$\scriptstyle s$}}}
    \raise2pt\hbox{$\scriptstyle \neg$}}}

\begin{document}

\title{
Lorentz and CPT tests with charge-to-mass ratio comparisons in Penning traps
} 

\author{Yunhua Ding and Mohammad Farhan Rawnak}

\affiliation{Department of Physics, Gettysburg College, Gettysburg, Pennsylvania 17325, USA}

\date{\today}

\begin{abstract}

Applications of the general theory of quantum electrodynamics
with Lorentz- and CPT-violating operators of mass dimensions up to six
are presented to Penning-trap experiments comparing charge-to-mass ratios 
between particles and antiparticles. 
Perturbation theory is used to derive Lorentz- and CPT-violating contributions 
to the energy levels and cyclotron frequencies 
of confined particles and antiparticles.
We show that whether the experimental {\it interpreted} quantity
$(|q|/m)_{\ol{w}}/(|q|/m)_{w} - 1$ is a clean measure of a CPT test 
depends on the context of the relevant theory.
Existing experimental results of charge-to-mass ratio comparisons 
are used to obtain first-time constraints on 69 coefficients for Lorentz and CPT violation. 

\end{abstract}

\maketitle

\section{Introduction}

Invariance under Lorentz transformations is one of 
the fundamental symmetries of both 
General Relativity and the Standard Model of particle physics.
However, 
tiny violations of Lorentz invariance could naturally emerge 
via spontaneous symmetry breaking 
in a fundamental theory unifying gravity with quantum physics,
such as string 
theory~\cite{ksp}.
In realistic effective field theory
any violation of CPT symmetry,
the invariance under the combined transformation of 
charge conjugation C, parity inversion P, and time reversal T, 
is accompanied by Lorentz 
violation~\cite{ck,owg}.
It follows that testing Lorentz symmetry includes CPT tests as well. 
Motivated by this,
many high-precision experiments
in various subfields of physics
have been performed to search for 
a variety of Lorentz- and CPT-violating 
signals~\cite{tables}.

Testing Lorentz and CPT symmetry requires either the study of 
the effects of a physical system under rotations or boosts,
or comparing the fundamental properties of a particle such as 
lifetime, charge-to-mass ratio, and $g$ factor
to these of its antiparticle.
Among the high-precision tests of Lorentz and CPT invariances,
the Penning trap is of particular interest,
as it provides a stable confinement of a particle or an antiparticle,
permitting highly precise measurements and comparisons of its properties. 
Meanwhile,
the Earth provides a natural rotating and boosting frame to study 
these properties under Lorentz transformations.
Impressive sensitivities have been achieved by Penning-trap experiments.
For example,
the proton and antiproton charge-to-mass ratios were compared to
parts per 
trillion~\cite{ul15}.
For the $g$ factors of electrons and positrons,
as well as these of protons and antiprotons,
the precision achieved at parts per 
billion~\cite{vd87, 17sm}.
The prospects of testing Lorentz and CPT symmetry in Penning-trap experiments
measuring the $g$ factors of particles and antiparticles were addressed in
Ref.~\cite{16dk}.
To extend that work,
we focus in this paper on searches for Lorentz and CPT violation
using the charge-to-mass ratio comparisons between particles and antiparticles
confined in a Penning trap.  

The comprehensive framework to study Lorentz and CPT violation 
in the context of effective field theory
is known as the Standard-Model Extension 
(SME)~\cite{ck,akgrav},
which is constructed from the action of General Relativity and the Standard Model
by adding all possible Lorentz-violating terms.
Each of these terms is formed from a coordinate-independent 
contraction of a Lorentz-violating operator with a corresponding controlling coefficient.
The subset of the SME containing operators of power-counting renormalizable 
mass dimension $d\leq 4$ is called the minimal SME,
while the nonminimal SME restricts attention to operators of mass dimensions $d>4$
and is assumed to produce higher-order corrections to conventional physics.

Both the minimal and nonminimal SME can produce various 
Lorentz- and CPT-violating effects in Penning-trap experiments,
including those measuring the $g$ factor 
and charge-to-mass ratio
of a confined particle or 
antiparticle~\cite{vd87,de99,17sc,17sm,ga99,ul15}. 
These effects include shifts in the cyclotron and anomaly frequencies
that can depend on sidereal time 
and also differ between particles and antiparticles.
The original theoretical work using the minimal SME 
to study Lorentz and CPT violation was given in
Ref.~\cite{bkr97,bkr98}.
It was recently extended to the nonminimal SME by including 
Lorentz- and CPT-violating operators of dimensions up to six,
together with applications to Penning-trap experiments comparing 
the $g$ factors between a particle and an 
antiparticle~\cite{16dk}.  
The searches for the effects arising from sidereal variations 
in these experiments have also been 
discussed~\cite{mi99,19ding,sm19}.

However, 
no treatment on the nonminimal SME effects in Penning-trap experiments
comparing charge-to-mass ratios exists in the literature to date. 
Addressing these nonminimal effects is of significance 
as it can reveal additional measurable Lorentz- and CPT-violating
signals due to the interactions of the particle or antiparticle 
with the electromagnetic fields in the trap.  
More generally,
studying the nonminimal SME sector can provide crucial insights
to many aspects of Lorentz and CPT violation,
such as noncommutative Lorentz-violating quantum 
electrodynamics~\cite{ha00,chklo01},
Lorentz-violating models in 
supersymmetry~\cite{susy},
or foundational issues including causality and 
stability~\cite{causality}
and the underlying pseudo-Riemann–Finsler 
geometry~\cite{finsler}.

In this work,
we address this gap by studying the nonminimal SME effects 
arising from particle and antiparticle charge-to-mass ratio measurements
in Penning traps. 
The theory of Lorentz- and CPT-violating electrodynamics with operators 
of mass dimensions up to 
six developed in 
Ref.~\cite{16dk} 
provides a partial guide to investigate these effects. 
Applying perturbation theory we derive the leading-order contributions 
due to Lorentz and CPT violation to cyclotron frequencies
and then relate them to charge-to-mass ratio comparisons.
We also address the question of whether a comparison 
of the experimental {\it interpreted} charge-to-mass ratios
between particles and antiparticles is a clean CPT test 
and conclude that it depends on the context of the relevant theory. 
Taking published results including the sidereal studies from Penning-trap experiments,
we extract first-time constraints on 69 SME coefficients.
The results obtained in this work are complementary to 
existing ones from comparisons of the $g$ factors
between particles and antiparticles in Penning-trap 
experiments~\cite{16dk,19ding},
the studies of the anomalous magnetic moment of muons confined in a storage 
ring~\cite{muon,gkv14},
the spectroscopic investigations of hydrogen, antihydrogen, and other related 
systems~\cite{kv15},
and clock-comparison 
experiments~\cite{kv18}.

This work is organized as follows.
In Section~\ref{theory},
we revisit the theory of quantum electrodynamics 
with Lorentz- and CPT-violating operators of mass dimensions up to six
and derive the perturbative Hamiltonian at leading order 
in Lorentz and CPT violation. 
We next turn in Section~\ref{application} the applications to Penning-trap experiments. 
The dominant energy shifts due to Lorentz and CPT violation of a confined particle
or antiparticle are given in subsection \ref{energy shifts},
followed in subsection~\ref{cyclotron frequency shifts}
by the corresponding cyclotron frequency shifts.
We address in subsection \ref{transformation under rotations} 
the general transformation of the coefficients for Lorentz violation 
between different frames.
This leads to a discussion in subsection \ref{experimental signals} 
of possible measurable Lorentz- and CPT-violating signals in Penning-trap experiments 
comparing charge-to-mass ratios between particles and antiparticles.
In subsection \ref{experimental sensitivities},
we use published experimental results 
to extract first-time constraints on 69 SME coefficients
and summarize them in
\tab{cons}.
Finally,
we given in Sec. \ref{summary} the summary of this work. 
Three appendices are given at the end of the paper 
for the reader' convenience.
In Appendix \ref{lag},
we reproduce the full Lagrange density of quantum electrodynamics
with Lorentz- and CPT-violating operators of mass dimensions $d\leq 6$.
The explicit calculation result of the perturbative energy shifts
is given in Appendix \ref{perturbative energy shifts},
followed in Appendix \ref{transformations}
by the transformation results for the related coefficients for Lorentz violation.

Throughout the paper,
we follow the notation used in Ref.~\cite{km13, 16dk},
unless otherwise specified. 
In particular, 
we adopt natural units with $\hbar =c = 1$
and express mass units in GeV.

\section{Theory}
\label{theory}

In this section,
we focus on the theory 
by revisiting the Lagrange density of Lorentz-violating spinor
electrodynamics with operators of mass dimensions up to 
six~\cite{16dk},
and deriving the related perturbative Hamiltonian at leading order 
in Lorentz and CPT violation.

\subsection{Lagrange density}

In the framework of the SME,
the general Lorentz-violating Lagrange density that preserves U(1) gauge invariance
for a single Dirac fermion field $\ps$ of charge $q$ and mass $\m$
coupled to an electromagnetic field $A_\mu$ is given by
\bea
\cl_\ps =
\cl_0 + \half \psb \Qhat \ps + {\rm H.c.} , 
\label{fermlag}
\eea
where $\cl_0 = \half \psb (\ga^\mu i D_\mu - \m ) \ps + {\rm H.c.}$ 
is the conventional Lorentz-invariant QED Lagrange density,
with $iD_\mu$ being the covariant derivative from the minimal coupling 
$iD_\mu \equiv i \prt_\mu - q A_\mu$
and H.c. denoting Hermitian conjugate. 
$\Qhat$ is a general $4\times4$ Lorentz-violating operator involving 
the covariant derivative $iD_\mu$ and the antisymmetric electromagnetic field tensor
$F_\mn \equiv \prt_\mu A_\nu - \prt_\nu A_\mu$.
From the hermiticity of the Lagrange density \rf{fermlag},
$\Qhat$ satisfies $\Qhat = \ga_0 \Qhat^\dag\ga_0$.
The spin content of $\Qhat$ can be shown by
expanding it in the basis of the 16 Dirac matrices,
\bea
\Qhat 
=
\Shat
+i\Phat \ga_5
+\Vhat^\mu \ga_\mu
+\Ahat^\mu \ga_5\ga_\mu
+\half \That^\mn \si_\mn ,
\label{qhatsplit}
\eea
where the 16 operators 
$\{\Shat,\Phat,\Vhat^\mu,\Ahat^\mu,\That^\mn\}$
are Dirac-scalar functions of mass dimension one
formed from the contraction of coefficients for Lorentz violation and 
operators including $iD_\mu$ and $F_\mn$ in general.
For example,
one of the dimension-five terms in $\Ahat^\mu$ takes the form $-\half  \bcf 5 \mab F_\ab$,
where $\bcf 5 \mab $ is the controlling coefficient for Lorentz violation.
As shown in Ref.~\cite{16dk},
this term can produce both Lorentz- and CPT-violating 
effects in experiments measuring the magnetic moment
of a particle or an antiparticle with a Penning trap. 

The explicit form of the Lagrange density \rf{fermlag}
at arbitrary mass dimension
in the free-fermion limit $A_\mu = 0$
has been studied in 
Ref.~\cite{km13}.
For a Dirac fermion interacting with a nonzero $A_\mu$,
its expression for mass dimension $d \leq 6$
was constructed in 
Ref.~\cite{16dk},
where a set of novel $F$-type coefficients for Lorentz violation 
associated with $F_\mn$ was discussed.
An extension to arbitrary mass dimension for the interacting case
was recently given by
Ref.~\cite{19kl}.
For other SME sectors, 
a similar analysis of the quadratic terms in the photon sector 
at arbitrary mass dimension has been presented in
Ref.~\cite{km09},
as well as extensions to the neutrino 
sector~\cite{km12},
and the gravity 
sector~\cite{nonmingrav}.
Since the Lagrange density constructed in
Ref.~\cite{16dk}
serves as the theoretical basis of the present work,
for convenience we reproduce these terms in Appendix \ref{lag}.

\subsection{Perturbative Hamiltonian}

Given no Lorentz- and CPT-violating signals have been
observed in experiments to date,
any such possible symmetry-violating effects must be tiny.
Therefore,
the contributions from the Lorentz-violating operator $\Qhat$ to the Hamiltonian 
related to the Lagrange density \rf{fermlag} 
can be treated as perturbative.
To derive the explicit expression of the perturbative Hamiltonian $\de\cH$,
we start from the modified Dirac equation in the momentum space,
\bea
\label{moddirac}
(p \cdot \ga - \m + \Qhat ) \ps = 0 ,
\eea
where $p_\mu \equiv  iD_\mu \equiv (i \prt_\mu - q A_\mu)$.
The exact Hamiltonian $\cH$ 
can then be defined from equation \rf{moddirac} via 
\bea
\cH  \ps \equiv p^0 \ps = \ga_0 (\pvec \cdot \gavec + \m - \Qhat) \ps ,
\eea
where $p^0$ is the exact energy of the physical system including Lorentz violation. 
Separating the exact Hamiltonian $\cH$ into the sum of 
the conventional Hamiltonian $\cH_0$ and the perturbative part $\de \cH$
due to Lorentz and CPT violation,
we identify the exact perturbative Hamiltonian to be
$\de\cH = -\ga_0 \Qhat$ .

It is challenging to construct the perturbative Hamiltonian $\de\cH$ directly 
since the Lorentz-violating operator $\Qhat$ in general contains powers of $p_0$ 
and thus includes the perturbative Hamiltonian $\cH$ itself.
This implies that the standard procedure cannot be adopted to obtain 
a Dirac Hamiltonian operator generating time translations on the wave function. 
For certain cases of $\Qhat$,
a field redefinition at the level of the Lagrange density 
can be performed to overcome this difficulty 
by removing the additional time 
derivatives~\cite{bkr98}. 
However, 
at the leading order in Lorentz violation,
a more general procedure,
first presented in 
Ref.~\cite{km12},
can be adopted by noticing that
any contributions to $\de\cH$ due to the exact Hamiltonian $\cH$ 
are at second or higher orders in Lorentz violation.
Therefore,
to obtain the leading-order effects,
the perturbative Hamiltonian $\de\cH$ can be evaluated 
using the unperturbative energy $E_0$ for $p^0$,
\bea
\label{pertbh}
\de\cH \approx -\ga_0 \Qhat |_{p^0 \to E_0} ,
\eea
where $E_0$ can be derived by solving the relevant conventional Dirac equation
for the physical system.

\section{Application to the Penning trap}
\label{application}

In this section,
we apply the above theory to experiments involving Penning traps.
Using perturbation theory we derive the energy shifts due to Lorentz
and CPT violation of particles and antiparticles confined in a Penning tap. 
Then we obtain the dominant shifts in their cyclotron frequencies,
followed by a discussion of general frame changes to
study the transformation under rotations.
This leads to investigations of possible measurable signals in Penning-trap 
experiments comparing the charge-to-mass ratios between particles and antiparticles,
including a discussion of the CPT test. 
By identifying the relations between experimental measured quantities
and coefficients for Lorentz violation,
we obtain first-time constraints on 69 SME coefficients from published Penning-trap results.

\subsection{Energy shifts}
\label{energy shifts}

For precision experiments involving particles or antiparticles 
confined in a Penning trap,
the relevant experimental observables of interest are frequencies,
which are the energy differences between energy levels.
To obtain the shifts in the energy levels of a confined particle
due to Lorentz and CPT violation,
we apply perturbation theory
\bea
\de E_{n, \pm}= \vev{\ch_{n, \pm}|\de\cH|\ch_{n, \pm}},
\label{deE}
\eea
where $\de \cH$ is the perturbative Hamiltonian given by \rf{pertbh},
$\ch_{n, \pm}$ denote the unperturbative four-component stationary eigenstates 
of level number $n$ and spin $\pm$ for the positive-energy fermion,
and $\de E_{n, \pm}$ are the corresponding perturbative energy shifts 
due to Lorentz and CPT violation.

In Penning-trap experiments,
the dominant effects in the unperturbative energy spectrum arise 
from the interaction of the confined particle or antiparticle 
with the constant magnetic field of the trap. 
The quadrupole electric field,
which varies with position to provide the axial confinement, 
generates weaker effects suppressed by a factor of $E/B$.
In a typical trap with $E \approx 20$~kV/m and $B \approx 5$~T,
this ratio is about $10^{-5}$ in natural units.
Therefore,
we start the theoretical analysis with an idealized scenario where 
a relativistic quantum fermion moves in a uniform magnetic field only. 
The unperturbative fermion eigenstates $\ch_{n, \pm}$
in the absence of Lorentz and CPT violation
can be obtained by solving the conventional Dirac equation 
with the minimal coupling for a spin-1/2 fermion of mass $m$
and charge $q\equiv \si |q|$ in a constant magnetic field. 
For calculation definiteness,
we fix the gauge with $A^\mu = (0,x_2 B,0,0) = (0,-x^2 B,0,0)$
so that the magnetic field is $\mbf B = B \hat x_3$, 
pointing the positive $x^3$ axis in the apparatus frame. 

After some calculation,
we present in 
Appendix~\ref{perturbative energy shifts} 
the explicit result of the energy shifts $\de \ens^w$ 
for a fermion of species $w$ and charge sign $\si$ in a magnetic field 
$\mbf B = B \hat x_3$ due to Lorentz-violating operators 
appearing in 
$\cl^{(3)}$, $\cl^{(4)}$, $\cl^{(5)}_D$, and $\cl^{(6)}_D$,
given in
Appendix \ref{lag}.
The additional energy shift contributions from operators 
in $\cl^{(5)}_F$ and $\cl^{(6)}_F$
can be obtained via substitutions listed in (40) in
Ref~\cite{16dk},
while terms in $\cl^{(6)}_{\prt F}$ produce no energy shift contributions 
as $\prt_\al F_{\be \ga}=0$ for a uniform magnetic field in a Penning trap.
In obtaining the result \rf{fullen},
we note that the axial motion of the confined particle or antiparticle 
in a Penning trap is purely induced by the electric field,
this means terms involving one or more powers of the Landau momentum $p_3$ 
appearing in the energy shift calculation are also 
suppressed by one or more powers of the ratio $E/B$.
Therefore,
we disregard such terms in result~\rf{fullen}
to obtain the leading-order contributions.
Note also that the unperturbed positive eigenenergies in result \rf{fullen}
take the form $E_{n, \pm 1}^w = \sqrt {m_w^2 + (2n+1 \mp \si)|qB|}$.

As shown in
Appendix~\ref{perturbative energy shifts},
the full energy shifts \rf{fullen}  
depend on several variables,
including the charge sign $\si$ of the particle,
the spin orientation,
and the level number~$n$.
The dependence on the direction of the magnetic field
is reflected by the indices of the coefficients for Lorentz violation
as the calculations are performed in the apparatus frame
with the magnetic field along the positive $x^3$ axis.
The magnitude dependence is evident from terms involving powers of $|qB|$.
Since in a typical Penning-trap experiment
a particle with $1e$ charge in a magnetic field of $B \approx 5$ T  
corresponds to $|qB| \approx 10^{-16}$~GeV$^2$ in natural units,
to obtain the leading-order contributions due to the magnetic field
we can expand terms containing $E_{n, \pm 1}^w$ in Taylor series of $|qB|$ 
and keep only up to the linear terms in $|qB|$ in the result. 

With the above approximations and including also the contributions from 
operators in $\cl^{(5)}_F$ and $\cl^{(6)}_F$,
we can rewrite the perturbative energy shifts \rf{fullen}
in the form
\bea
\label{linear-en}
&&
\de E_{n, \pm 1}^w 
\nn \\
&=&
\atw w 0 
\mp \si \btw w 3
- \mftw w 3 B
\pm \si \bftw w {33} B 
\nn\\
&&
\hskip -3pt
+ \big(
\pm \si \bptw w {3} 
- m_w [\ctw w {00} + (\ctw w {11} + \ctw w {22})_{s}]
\big)
\fr{(2n+1\mp \si)|qB|}{2 m_w^2}
\nn\\
&&
\hskip -3pt
+ 
\big(
\mp \si (\btw w {311} + \btw w {322})
- \fr{1}{m_w} (\ctw w {11} + \ctw w {22})_{\vos}
\big) 
\fr{(2n+1)|qB|}{2},
\nn \\
\eea
where the various tilde coefficients are defined by 
\bea
\label{tild}
\atw w 0
&=&
\acmw 3 0 w
- \mw \ccmw 4 {00} w
- \mw \ecmw 4 0 w
+ \mw^2 \mcw 5 {00} w
\nn \\
&&
+ \mw^2 \acw 5 {000} w
- \mw^3 \ccw 6 {0000} w
- \mw^3 \ecw 6 {000} w,
\nn\\
\btw w 3
&=&
\bcmw 3 3 w
+ \Hcmw 3 {12} w
- \mw \dcmw 4 {30} w
- \mw \gcmw 4 {120} w
+ \mw^2 \bcw 5 {300} w
\nn \\
&&
+ \mw^2 \Hcw 5 {1200} w
- \mw^3 \dcw 6 {3000} w
- \mw^3 \gcw 6 {12000} w,
\nn\\
\mftw w 3
&=&
\mcfw 5 {12} w
+ \acfw 5 {012} w
- \mw \ccfw 6 {0012} w
- \mw \ecfw 6 {012} w,
\nn\\
\bftw w {33}
&=&
\bcfw 5 {312} w
+ \Hcfw 5 {1212} w
- \mw \dcfw 6 {3012} w
- \mw \gcfw 6 {12012} w ,
\nn\\
\bptw w {3}
&=&
b_{w}^{3} + m_w (g_{w}^{120} - g_{w}^{012} + g_{w}^{021}) 
- m_w^2 b_{w}^{(5)300} 
\nn \\
&&
- 2 m_w^2 (H_{w}^{(5)1200} - H_{w}^{(5)0102} + H_{w}^{(5)0201})
\nn \\
&&
+ 2 m_{w}^3 d_{w}^{(6)3000}  
\nn \\
&&
+ 3 m_w^3 (g_{w}^{(6)12000}-g_{w}^{(6)01002}+g_{w}^{(6)02001}) ,
\nn\\
\ctw w {00}
&=&
c_{w}^{00} - m_w m_{w}^{(5)00} - 2 m_w a_{w}^{(5)000} 
\nn \\
&&
+ 3 m_w^2 c_{w}^{(6)0000} + 2 m_w^2 e_{w}^{(6)000},
\eea
and the ``11+22" types of tilde coefficients are defined by
\bea 
\label{jj}
(\ctw w {jj})_{s}
& = &
c_{w}^{jj} 
- 2 m_w a_{w}^{(5)j0j} 
+ 3 m_w^2 c_{w}^{(6)j00j}  ,
\nn\\ 
(\ctw w {jj} )_{\vos}
& = &
- m_w a_{w}^{(5)0jj} 
- m_w m_{w}^{(5)jj} 
\nn \\
&&
+ 3 m_w^2 c_{w}^{(6)00jj} 
+ 3 m_w^2 e_{w}^{(6)0jj} ,
\nn\\
\btw w {3jj} 
& = &
b_{w}^{(5)3jj} + H_{w}^{(5)12jj} 
\nn \\
&&
- 3 m_{w} d_{w}^{(6)30jj} 
- 3 m_{w} g_{w}^{(6)120jj} ,
\eea
with $j$ taking values of 1 and 2 only.
The subscripts $s$ and 
$\mathrel{\rlap{\lower0pt\hbox{\hskip0.5pt{$s$}}}\raise2pt\hbox{$\neg$}}$
in the above $\ctw w {jj}$ tilde coefficients
specify the fact that $(\ctw w {jj})_{s}$ produce both spin-independent
and spin-dependent energy shift contributions,
while $(\ctw w {jj} )_{\vos}$ give only spin-independent ones,
which is evident from the corresponding proportional factors 
$2n+1\mp \si$ and $2n+1$ in result \rf{linear-en}.
We note that the energy shift contributions from
tilde coefficients 
$\atw w 0$,
$\btw w 3$,
$\mftw w 3$ ,
and $\bftw w {33}$
are independent of the level number $n$.
Expression \rf{linear-en} extends the energy shift result obtained in 
Ref.~\cite{16dk}
by including terms linear in $|qB|$.
These terms can lead to nonzero contributions to the cyclotron frequencies,
as will be shown in the next subsection.

The corresponding shifts in the antifermion energy levels due to Lorentz and CPT violation
are given by
\bea
\label{deEs}
\de E^c_{n,\pm}= \vev{\ch^c_{n,\pm}|\de\cH^c |\ch^c_{n,\pm}} ,
\eea
where $\ch^c_{n,\pm}$ are the corresponding positive-energy antifermion eigenstates,
which can be derived from the negative-energy fermion solutions $\ch_{n,\pm}$
via charge conjugation in the usual way,
and $\de\cH^c$ is the perturbative Hamiltonian for the antifermion,
which can also be obtained from $\de\cH$ in a similar way.
Applying Eq. \rf{deEs},
the expression for the perturbative energy shifts of the corresponding antifermion
is found to have the same form as that of a fermion,
except that the spin is reversed and 
the contributions are controlled by a set of starred tilde quantities, 
\begin{flalign}
\label{linear-ens}
&
\de E_{n, \pm 1}^{\ol{w}} 
\nn &\\
 = &
\ - \atws w 0 
\pm \si \btws w 3
- \mftws w 3 B
\mp \si \bftws w {33} B 
\nn &\\
&
+ \big(
\mp \si \bptws w {3} 
- m_w [\ctw w {00} + (\ctw w {11} + \ctw w {22})_{s}]
\big)
\fr{(2n+1\mp \si)|qB|}{2 m_w^2}
\nn &\\
&
+ 
\big(
\pm \si (\btws w {311} + \btws w {322})
- \fr{1}{m_w} (\ctw w {11} + \ctw w {22})_{\vos}
\big) 
\fr{(2n+1)|qB|}{2},
\nn &\\
\end{flalign}
where the starred tilde quantities are defined by 
\bea
\label{tilds}
\atws w 0
&=&
\acmw 3 0 w
+ \mw \ccmw 4 {00} w
- \mw \ecmw 4 0 w
- \mw^2 \mcw 5 {00} w
\nn \\
&&
+ \mw^2 \acw 5 {000} w
+ \mw^3 \ccw 6 {0000} w
- \mw^3 \ecw 6 {000} w,
\nn\\
\btws w 3
&=&
\bcmw 3 3 w
- \Hcmw 3 {12} w
+ \mw \dcmw 4 {30} w
- \mw \gcmw 4 {120} w
+ \mw^2 \bcw 5 {300} w
\nn \\
&&
- \mw^2 \Hcw 5 {1200} w
+ \mw^3 \dcw 6 {3000} w
- \mw^3 \gcw 6 {12000} w,
\nn\\
\mftws w 3
&=&
\mcfw 5 {12} w
- \acfw 5 {012} w
- \mw \ccfw 6 {0012} w
+ \mw \ecfw 6 {012} w,
\nn\\
\bftws w {33}
&=&
\bcfw 5 {312} w
- \Hcfw 5 {1212} w
+ \mw \dcfw 6 {3012} w
- \mw \gcfw 6 {12012} w ,
\nn\\
\bptw w {3}
&=&
b_{w}^{3} + m_w (g_{w}^{120} - g_{w}^{012} + g_{w}^{021}) 
- m_w^2 b_{w}^{(5)300} 
\nn \\
&&
+ 2 m_w^2 (H_{w}^{(5)1200} - H_{w}^{(5)0102} + H_{w}^{(5)0201})
\nn \\
&&
- 2 m_{w}^3 d_{w}^{(6)3000}  
\nn \\
&&
+ 3 m_w^3 (g_{w}^{(6)12000}-g_{w}^{(6)01002}+g_{w}^{(6)02001}) ,
\nn\\
\ctws w {00}
&=&
c_{w}^{00} - m_w m_{w}^{(5)00} + 2 m_w a_{w}^{(5)000} 
\nn \\
&&
+ 3 m_w^2 c_{w}^{(6)0000} - 2 m_w^2 e_{w}^{(6)000},
\eea
and the corresponding ``11+22" types of starred tilde coefficients are given by
\bea 
\label{jjs}
(\ctws w {jj})_{s}
& = &
c_{w}^{jj} 
+ 2 m_w a_{w}^{(5)j0j} 
+ 3 m_w^2 c_{w}^{(6)j00j}  ,
\nn\\ 
(\ctws w {jj} )_{\vos}
& = &
m_w a_{w}^{(5)0jj} 
- m_w m_{w}^{(5)jj} 
\nn \\
&&
+ 3 m_w^2 c_{w}^{(6)00jj} 
- 3 m_w^2 e_{w}^{(6)0jj} ,
\nn\\
\btws w {3jj} 
& = &
b_{w}^{(5)3jj} - H_{w}^{(5)12jj} 
\nn \\
&&
+ 3 m_{w} d_{w}^{(6)30jj} 
- 3 m_{w} g_{w}^{(6)120jj} .
\eea
In result~\rf{linear-ens},
the charge sign $\si$ of the antifermion is understood to change.
Comparing the result~\rf{linear-ens} to \rf{linear-en},
together with the relevant definitions 
\rf{tild}, \rf{jj}, \rf{tilds}, and \rf{jjs}, 
$\de E_{n, \pm 1}^{\ol{w}}$ can also be obtained from $\de E_{n, \pm 1}^{w}$
by reversing the charge sign,
the spin orientation,
and the signs of all CPT-odd coefficients.

We remark in passing that 
the indices of the tilde coefficients
appearing in results \rf{linear-en} and \rf{linear-ens}
and are defined in
\rf{tild}, \rf{jj}, \rf{tilds}, and \rf{jjs}
correctly represent their rotation properties.
For example,
the index pair ``12" on the right sides of these definitions 
is antisymmetric~\cite{anti},
which means it transforms like a single ``3" index under spatial rotations.
Coefficients with index ``0" or index pair ``00" are invariant under spatial rotations. 
The dependence of results \rf{linear-en} and \rf{linear-ens}
on only the index ``0", ``3", and ``11+22"
correctly reflects the cylindrical symmetry of the Penning trap.

\subsection{Cyclotron frequency shifts}
\label{cyclotron frequency shifts}

One of the key frequencies in a Penning-trap experiment
is the cyclotron frequency,
which is related to the charge-to-mass ratio 
of the confined particle or antiparticle.
The cyclotron frequency is defined as the energy difference 
between the $n=1$ and $n=0$ Landau levels. 
For example,
for particles $w=e^-$ and $p$,
the cyclotron frequencies in natural units are defined as
\bea
\label{wc}
\omega_c^{e^-} \equiv E_{1, -1}^{e^-} -  E_{0,  -1}^{e^-} ,  
\hskip 7pt 
\omega_c^{p} \equiv E_{1, +1}^{p} -  E_{0,  +1}^{p} ,
\eea 
respectively.
For the corresponding antiparticles $\ol{w}=e^+$ and $\bar{p}$,
the cyclotron frequencies are defined in a similar way,
except the spin directions are reversed,
\bea
\label{wcs}
\omega_c^{e^+} \equiv E_{1, +1}^{e^+} -  E_{0,  +1}^{e^+} ,  
\hskip 7pt 
\omega_c^{\bar{p}} \equiv E_{1, -1}^{\bar{p}} -  E_{0,  -1}^{\bar{p}} .
\eea 
In the presence of Lorentz and CPT violation,
the perturbative energy shifts \rf{linear-en} and \rf{linear-ens} can lead to corrections 
to these cyclotron frequencies.
Using the definitions \rf{wc} together with the result \rf{linear-en},
we find the shifts in the cyclotron frequencies for electrons $w=e^-$
and protons $w=p$
have the same expression, 
\beq
\label{wcshift}
\fr{\de \omega_c^{w}}{eB} 
=
\dfrac{1}{m_w^2} \bptw w {3} 
- \dfrac{1}{m_w} (\ctw w {00} + \ctw w {11} + \ctw w {22})
- (\btw w {311} + \btw w {322}) ,
\eeq
where the tilde coefficients $\ctw w {jj}$ with $j=1$ or $2$ 
are the sum of the two pieces
defined in \rf{jj},
\bea
\ctw w {jj} 
&=&
(\ctw w {jj})_{s} + (\ctw w {jj} )_{\vos}
\nn\\
&=&
c_{w}^{jj} 
- 2 m_w a_{w}^{(5)j0j} 
+ 3 m_w^2  c_{w}^{(6)j00j} 
- m_w a_{w}^{(5)0jj} 
\nn\\
&&
- m_w m_{w}^{(5)jj} 
+ 3 m_w^2 c_{w}^{(6)00jj} 
+ 3 m_w^2 e_{w}^{(6)0jj}  .
\eea
 
The above result \rf{wcshift} shows that the shifts in the cyclotron frequencies 
due to Lorentz and CPT violation
depend only on three tilde quantities
$\bptw w {3}$,  
$\ctw w {00} + \ctw w {11} + \ctw w {22}$,
and
$\btw w {311} + \btw w {322}$
in the apparatus frame,
among which the piece $\ctw w {00}$ is invariant under rotations 
but breaks Lorentz symmetry under boosts,
while the others violate Lorentz symmetry under both rotations and boosts. 
All of tilde quantities involve a mixture of CPT-even and CPT-odd coefficients.
No $F$-type coefficients for Lorentz violation 
appear in result \rf{wcshift}
as they produce energy shift contributions that are 
independent of the Landau level number $n$,
as evident from the result \rf{linear-en},
and hence are unobservable in the cyclotron frequency shifts.

The corresponding cyclotron frequency shifts for antiparticles 
$\ol{w}=e^+$ and $\bar{p}$
can be obtained from \rf{wcshift} by replacing the usual tilde quantities 
by the corresponding starred ones,
given by
\beq
\label{wcsshift}
\fr{\de \omega_c^{\ol{w}}}{eB} 
=
- \dfrac{1}{m_w^2} \bptws w {3} 
- \dfrac{1}{m_w} (\ctws w {00} + \ctws w {11} + \ctws w {22})
+ (\btws w {311} + \btws w {322}) ,
\\
\eeq
where the starred tilde coefficients $\ctws w {jj}$ with $j=1$ or $2$ 
are defined by
\bea
\ctws w {jj} 
&=&
(\ctws w {jj})_{s} + (\ctws w {jj} )_{\vos}
\nn\\
&=&
c_{w}^{jj} 
+ 2 m_w a_{w}^{(5)j0j} 
+ 3 m_w^2  c_{w}^{(6)j00j} 
+ m_w a_{w}^{(5)0jj} 
\nn\\
&&
- m_w m_{w}^{(5)jj} 
+ 3 m_w^2 c_{w}^{(6)00jj} 
- 3 m_w^2 e_{w}^{(6)0jj}  .
\eea
It is observed that the only difference between
the particle result \rf{wcshift} and the antiparticle result \rf{wcsshift} 
is the sign of all the CPT-odd coefficients,
as expected. 

\subsection{Transformation under rotations}
\label{transformation under rotations}

The cyclotron frequency shifts  \rf{wcshift} and \rf{wcsshift} 
are expressed in the apparatus frame
with the positive $\hat x_3$ axis chosen to be aligned 
with the magnetic field in the trap.
However, 
this frame is noninertial due to the Earth's rotation.
To compare results from different experiments searching for 
Lorentz and CPT violation,
a standard canonical frame is adopted in the literature 
which is called the Sun-centered 
frame~\cite{sunframe0,sunframe}.
In this frame the cartesian coordinates are labeled by $X^J\equiv (X,Y,Z)$,
with the $Z$ axis aligned along the Earth's rotation axis,
the $X$ axis pointing from the Earth to the Sun, 
and the time $T$ chosen to have origin at the vernal equinox 2000.
The coefficients for Lorentz violation in this frame 
can be assumed to be constants in time and 
space~\cite{ck,akgrav}.

To relate the coefficients for Lorentz violation
from the Sun-centered frame to the apparatus frame,
we introduce a third frame called the standard laboratory frame
$x^j \equiv (x,y,z)$,
in which the $z$ axis points to the local zenith,
the $x$ axis is aligned with the local south,
and the $y$ axis completes a right-handed coordinate system. 
The convenient choice of the positive $\hat x_3$ axis of the apparatus frame
as the direction of the magnetic field of the trap 
may result in a nonzero angle to the $\hat z$ axis of the standard laboratory frame.
Therefore,
to relate the coordinates from the apparatus frame to the Sun-centered frame,
we define two rotation matrices
$R^{aj}$ and $R^{jJ}$,
with $R^{aj}$ relating the standard laboratory frame $x^j\equiv (x,y,z)$
to the apparatus frame $x^a\equiv (x^1,x^2,x^3)$ by $x^a = R^{aj} x^j$, 
and $R^{jJ}$ connecting
$X^J\equiv (X, Y, Z)$ of the Sun-centered frame
to $(x,y,z)$ of the standard laboratory frame by $x^{j} = R^{jJ} X^J$.
The expression of $R^{jJ}$ is given 
by~\cite{sunframe0,sunframe}
\beq
R^{jJ}=\left(
\begin{array}{ccc}
\cos\ch\cos\om_\oplus T_\oplus
&
\cos\ch\sin\om_\oplus T_\oplus
&
-\sin\ch
\\
-\sin\om_\oplus T_\oplus
&
\cos\om_\oplus T_\oplus
&
0
\\
\sin\ch\cos\om_\oplus T_\oplus
&
\sin\ch\sin\om_\oplus T_\oplus
&
\cos\ch
\end{array}
\right),
\label{rot}
\eeq
where $\om_\oplus \simeq 2\pi/(23{\rm ~h} ~56{\rm ~min})$
denotes the sidereal frequency of the Earth's rotation,
$T_\oplus$ specifies the local sidereal time,
and $\ch$ is the colatitude of the laboratory.
The rotation $R^{aj}$ can be specified in general 
by a suitable set of Euler angles $(\al, \be, \ga)$.
Adopting the convenient ``$y$-convention" of the 
rotation~\cite{ycon},
$R^{aj}$ is found to have the form
\bea
R^{aj} &=&
\left(
\begin{array}{ccc}
\cos\ga &\sin\ga  &0\\
-\sin\ga &\cos\ga &0\\
0 &0 &1\\
\end{array} 
\right)
\times
\left(
\begin{array}{ccc}
\cos\be &0 &-\sin\be\\
0 &1 &0\\
\sin\be &0 &\cos\be\\
\end{array} 
\right)
\nn \\
&&
\times
\left(
\begin{array}{ccc}
\cos\al &\sin\al &0\\
-\sin\al &\cos\al &0\\
0 &0 &1\\
\end{array} 
\right).
\label{euler}
\eea

Putting the above discussion together,
the relation between the coordinates of the apparatus frame
and these of the Sun-centered frame 
can be obtained by the following expression,
\bea
\label{axis}
x^{a} = R^{aj} x^j = R^{aj} R^{jJ} x^J ,
\eea
which can be used to relate the coefficients for Lorentz violation
in these two frames. 
In the special case where the magnetic field is vertical upward,
which means axes $\hat x_3$ and $\hat z$ are in the same direction,
the Euler angles become $(\al, \be, \ga)=(0,0,0)$ and $R^{aj}$
reduces to the identity matrix.

The rotation matrix \rf{rot} reveals the dependence on 
the sidereal time and laboratory geometric location 
of the coefficients for Lorentz violation observed in the apparatus frame.
As a result,
experimental observables for Lorentz violation can oscillate 
at harmonics of the sidereal frequency $\om_\oplus$ of the Earth's rotation,
and have different expressions for laboratories at different colatitudes.
To explicitly illustrate this,
we consider a Penning-trap experiment located at colatitude $\ch$ 
with a magnetic field along the $\hat{z}$ axis.
The tilde coefficients for Lorentz violation in
the cyclotron frequency shifts \rf{wcshift} and \rf{wcsshift}
that break Lorentz symmetry under rotations are
$\bptw w {3}$, 
$\ctw w {11} + \ctw w {22}$,
$\btw w {311} + \btw w {322}$,
$\bptws w {3}$, 
$\ctws w {11} + \ctws w {22}$,
and
$\btws w {311} + \btws w {322}$
in the apparatus frame. 
To express the sidereal-time and geometric dependence,
we take tilde quantity $\ctw w {11} + \ctw w {22}$ as an example.
Applying the rotation \rf{axis} with $R^{aj}$ being the identity matrix yields
\bea
\label{trans-ex}
\ctw w {11} + \ctw w {22} 
&=& 
\ctodt  
\big(  - \half (\ctw w {XX} -  \ctw w {YY}) \sin^2 \ch \big)
\nn \\
&&
+ \stodt
(  -  \ctw w {(XY)} \sin^2 \ch)
\nn \\
&& 
+  \codt  
(  - \ctw w {(XZ)}  \sin 2\ch )
\nn \\
&&
+ \sodt
(  - \ctw w {(YZ)}  \sin 2\ch  )
\nn \\
&&
+ 
\frac{1}{4} (\ctw w {XX} + \ctw w {YY})(3+\cos 2\ch) + \ctw w {ZZ} \sin^2\ch ,
\nn \\
\eea 
where parentheses on two indices of the coefficients
imply symmetrization with a factor of 1/2.
For instance, 
$\ctw w {(XY)} = (\ctw w {XY} + \ctw w {YX} )/2$.
Result \rf{trans-ex} shows that the tilde quantity $\ctw w {11} + \ctw w {22}$
can be expressed in terms of the six independent quantities 
$\ctw w {(JK)}$ with $J,K=X,Y,Z$ in the Sun-centered frame,
producing up to second harmonics in the sidereal frequency of the Earth's rotation. 
The colatitude dependence is shown by the factors appearing in the 
amplitudes of the harmonic oscillations. 

If magnetic fields in different directions are used in a Penning trap, 
the tilde quantities
$\bptw w {3}$, 
$\ctw w {11} + \ctw w {22}$,
$\btw w {311} + \btw w {322}$,
$\bptws w {3}$, 
$\ctws w {11} + \ctws w {22}$,
and
$\btws w {311} + \btws w {322}$
in the apparatus frame can have different transformations
into the Sun-centered frame. 
In a typical Penning-trap experiment,
the magnetic field is oriented either horizontally or vertically. 
For analysis reference,
we present in 
Appendix~\ref{transformations}
the explicit transformation results of these tilde quantities
for the above two field orientations. 
These transformation results 
show that for a given fermion of species $w$ 
there are a total of 27 independent tilde coefficients
$\bptw w {J}$,
$\ctw w {(JK)}$,
and 
$\btw w {J(KL)}$
in the Sun-centered frame 
that are related to the cyclotron frequency shifts
in a Penning-trap experiment.
Taking into account the corresponding antifermion $\ol{w}$,
an additional 27 independent components can be accessed via
$\bptws w {J}$,
$\ctws w {(JK)}$,
and 
$\btws w {J(KL)}$. 
 
A different type of time variation of the coefficients for Lorentz violation
can arise from the revolution of the Earth about the Sun. 
This includes the effects from the boost $\be_\oplus \approx 10^{-4}$
of the Earth relative to the Sun,
and the boost $\be_L \approx 10^{-6}$ of the laboratory due to the Earth's rotation. 
As studied in the 
literature~\cite{gkv14,kv15,ca04,he08,space},
these effects are suppressed by one or more powers of these boost factors
compared to these from rotations,
and hence can be treated as negligible in the present work.

\subsection{Experimental signals}
\label{experimental signals}

The SME can produce various 
Lorentz- and CPT-violating effects in Penning-trap experiments
involving confined particles or antiparticles. 
One type of observable signal arises from the time variation 
of the experimental quantity measured in the laboratory frame.
This is because the magnetic field used in the trap sets up a set of instantaneous 
coordinates of the laboratory frame,
which rotates due to the Earth's rotation
and hence produces sidereal variations of the measured signals,
as discussed above in 
subsection~\ref{transformation under rotations}.
Performing a sidereal-variation analysis of the experimental data would
permit the study of the constant coefficients in the Sun-centered frame
that are related to different harmonic terms in the transformation. 
For example, 
from results \rf{wcshift} and \rf{wcsshift},
together with the related transformations presented in 
Appendix~\ref{transformations},
a sidereal-variation analysis of the cyclotron frequency shifts 
$\de \om_c^w$ and $\de \om_c^{\ol{w}}$
of the confined particles and antiparticles
would offer sensitivities to components of the tilde coefficients 
$\bptw w {J}$, 
$\bptws w {J}$, 
$\ctw w {(JK)}$,
$\ctws w {(JK)}$,
$\btw w {J(KL)}$
and 
$\btws w {J(KL)}$.

Another kind of Lorentz- and CPT-violating effect 
appears in comparative measurements  
between particles and antiparticles,
as the frequency shifts due to Lorentz and CPT violation 
could differ between particles and antiparticles.
For example, 
according to results \rf{wcshift} and~\rf{wcsshift},
the cyclotron frequencies
for a particle and its antiparticle are shifted differently,
with the contributions controlled by two different sets of tilde quantities. 
In a Penning trap experiment using a same magnetic field,
the only difference between 
$\de \om_c^w$ and $\de \om_c^{\ol{w}}$
is the sign for the CPT-odd coefficients,
so a comparison between these two frequency shifts 
$\De \om_c^w = \de \om_c^w - \de \om_c^{\ol{w}}$ 
would permit the cancellations of all the CPT-even effects,
making it a clean test of CPT symmetry. 
Together with the sidereal-variation analysis discussed above,
different components of these CPT-odd tilde coefficients 
in results \rf{wcshift} and \rf{wcsshift} can be extracted in principle.

In this work,
we discuss the above two types of signals in Penning-trap experiments 
involving measurements of the charge-to-mass ratios 
of a particle and an antiparticle. 
In a Lorentz-invariant scenario,
conventional quantum electrodynamics predicts that
the charge-to-mass ratio of a particle or an antiparticle
is related to its cyclotron frequency by
\bea
\label{qm}
\dfrac{|q|}{m}   = \dfrac{\om_c}{B} .
\eea  
By measuring the cyclotron frequency of a particle or an antiparticle
in a known magnetic field,
its charge-to-mass ratio is then determined. 
Note that in Lorentz-invariant quantum field theory,
the definitions of the charge and mass 
for a particle or antiparticle are based on
the coupling constants charactering the related interaction strength. 
Therefore,
the charge-to-mass ratio is an intrinsic property of a particle 
or antiparticle
which does not vary by the local experimental conditions,
such as the field configuration in the trap
or the location of the laboratory. 
Note also that both the charge and mass are Lorentz scalars 
in quantum field theory and are invariant 
under Lorentz transformation.
It follows that the charge, mass and the resulting charge-to-mass ratio 
defined in the context of Lorentz-invariant quantum field theory
are unchanged even though Lorentz symmetry is broken.  

For comparative measurements,
suppose a Penning-trap experiment uses 
a same magnetic field to measure the cyclotron frequencies
of a particle $w$ and its corresponding antiparticle $\ol{w}$ simultaneously.
The result can then be related to the comparison 
of the charge-to-mass ratios between a particle and its antiparticle, 
\bea
\label{ratio-li}
\dfrac{(|q|/m)_{\ol{w}}}{(|q|/m)_{w}} - 1
=
\dfrac{\om_c^{\ol{w}}}{\om_c^{w}} - 1 .
\eea 
In the Lorentz- and CPT-invariant scenario,
this difference is identically zero by the CPT theorem.
Therefore,
in the context of Lorentz-invariant quantum field theory 
(which also implies CPT invariance),
the experimental {\it interpreted} quantity
$(|q|/m)_{\ol{w}}/(|q|/m)_{w} - 1 $
is a clean measure of a CPT test. 

However,
in the presence of Lorentz violation,
the cyclotron frequency of a particle or antiparticle is shifted by
\rf{wcshift} or \rf{wcsshift},
respectively.  
This implies that the measured cyclotron frequency
becomes an experiment-dependent quantity,
as a function of the local sidereal time,
the colatitude of the laboratory,
and both the direction and magnitude of the magnetic field used in the trap.  
As a result,
the difference \rf{ratio-li} does not vanish in general 
and become an experiment-dependent quantity,
given by
\bea
\label{ratio-lv}
\dfrac{(|q|/m)_{\ol{w}}}{(|q|/m)_{w}} - 1
\longleftrightarrow
\dfrac{\om_c^{\ol{w}}}{\om_c^{w}} - 1
=
\dfrac{\de \om_c^{\ol{w}} -  \de \om_c^{w}} {\om_c^{w}} ,
\eea
where $\longleftrightarrow$ means the charge-to-mass ratio comparison
reported by the experiments are obtained by interpreting the measured difference 
$\om_c^{\ol{w}}/\om_c^{w} - 1$,
as the relation~\rf{qm} becomes an approximation 
in the presence of Lorentz violation. 
On the right side of Eq.~\rf{ratio-lv}, 
the Lorentz- and CPT-invariant pieces 
in the measured cyclotron frequencies are exactly cancelled
by the CPT theorem if a same magnetic field is used.
From the cyclotron frequency shifts \rf{wcshift} and \rf{wcsshift},
together with the transformation results in 
Appendix~\ref{transformations},
the difference $\de \om_c^{w} - \de \om_c^{w}$ on the right side of 
the equation \rf{ratio-lv} contains only CPT-odd coefficients for Lorentz violation,
producing pure CPT-violating effects.
In a general case where a comparison is made 
by using different magnetic fields, 
Eq. \rf{ratio-lv} becomes 
\beq
\label{ratio-lv-b}
\dfrac{(|q|/m)_{\ol{w}}}{(|q|/m)_{w}} - 1
\longleftrightarrow
\dfrac{\om_c^{\ol{w}}/B^*}{\om_c^{w}/B} - 1
=
\dfrac{\de \om_c^{\ol{w}}/B^* -  \de \om_c^{w}/B} {\om_c^{w}/B} ,
\eeq
where $B^*$ and $B$ are the strengths of the magnetic fields 
used for measuring $\om_c^{\ol{w}}$ and $\om_c^{w}$,
respectively.
It's clear that the CPT-even coefficients on the right side of Eq.~\rf{ratio-lv-b}
don't exactly cancel out due to the different magnetic strengths,
even when the magnetic fields are in the same direction. 
Therefore,
in this case the experimental {\it interpreted} quantity
$(|q|/m)_{\ol{w}}/(|q|/m)_{w} - 1$ is not a clean measure of a CPT test. 

Now we conclude that whether the experimental {\it interpreted} quantity
$(|q|/m)_{\ol{w}}/(|q|/m)_{w} - 1$ is a clean test of CPT symmetry
depends on the context of the relevant theory. 
In a Lorentz-invariant quantum field theory,
it can be used as a clean test of CPT symmetry.
However,
in a general Lorentz-violating scenario,
it is a clean measure of a CPT test only if 
a same magnetic field is used.  
Similar discussions for the $g$ factor comparisons
have also been addressed in Section IIIB1 in 
Ref.~\cite{16dk}.
The key point is that the experimental quantity
$(|q|/m)_{\ol{w}}/(|q|/m)_{w} - 1$ is obtained by interpreting 
the measured difference $\om_c^{\ol{w}}/\om_c^{w} - 1$
in the context of conventional quantum electrodynamics.
Lorentz violation does not modify the theoretical value of the charge-to-mass ratio
for a particle or antiparticle,
which is defined via conventional quantum electrodynamics.
What it affects are the measured cyclotron frequencies 
that are used by experiments to interpret the charge-to-mass ratios
and their comparison between a particle and an antiparticle.

At the end of this subsection,
we discuss a subtlety arising from Penning-trap experiments 
comparing the charge-to-mass ratios 
between an antiproton and a proton.
As the two particles have opposite charges,
the measurements of their cyclotron frequencies using the same trap
requires the reversal of the quadruple electric field.
To facilitate the experiment by eliminating the systematic
shifts caused by polarity switching of the trapping voltages,
a hydrogen ion (H$^-$) is used as a proxy for the proton.
This allows relatively fast exchange between hydrogen ions and antiprotons.
The comparison of the charge-to-mass ratios between an antiproton and a proton
can be related to that between an antiproton and a hydrogen ion by
\beq
\label{ratioH1}
\dfrac{(|q|/m)_{\bar{p}}}{(|q|/m)_{p}} - 1
=
\dfrac{(|q|/m)_{\bar{p}}}{R (|q|/m)_{\rm{H}^-}} - 1
\longleftrightarrow
\dfrac{\de \om_c^{\bar{p}} - R \de \om_c^{\rm{H}^-}} {R \om_c^{\rm{H}^-}} ,
\eeq
where $R=m_{\rm{H}^-}/m_p = 1.001089218754$
is the mass ratio of a hydrogen ion and a 
proton~\cite{ul15},
$\om_c^{\rm{H}^-}$ and $\de \om_c^{\rm{H}^-}$ are the cyclotron
frequency and the corresponding shifts for the hydrogen ion,
respectively. 

The cyclotron frequency shift $\de \om_c^{\rm{H}^-}$
for a hydrogen ion in the above result
can be obtained by taking $w=\rm{H}^-$ in expression~\rf{wcshift}.
The related coefficients for Lorentz violation 
become the effective ones for a hydrogen ion.
In the framework of the SME,
effective coefficients for a composite particle can be expressed 
in terms of the corresponding fundamental coefficients 
for Lorentz violation for its constituents.
In our case,
the related fundamental coefficients are these for electrons and protons. 
Deriving the exact relations between these coefficients can be challenging 
due to nonperturbative issues involving binding effects for the composite particle. 
However, 
a good approximation to these relations can be found by 
taking the lowest-order perturbation theory
and treating the hydrogen ion wave function as a product of 
the wave functions of a proton and two electrons.  
Ignoring the binding energies,
the energy shifts of the hydrogen ion due to Lorentz and CPT violation
can be approximated as the sum of these for its constituents.
The corresponding approximated relation of the shifts in their cyclotron frequencies 
can be found as 
\bea
\label{rela}
\de \om_{c}^{\rm{H}^-} \approx \de \om_{c}^{p} + 2 \de \om_{c}^{e^-} .
\eea
Substituting relation \rf{rela} into the result \rf{ratioH1} yields
\bea
\label{ratioH2}
\dfrac{(|q|/m)_{\bar{p}}}{(|q|/m)_{p}} - 1
\approx
\dfrac{\de \om_c^{\bar{p}} - R \de \om_c^{p} -2R  \de \om_c^{e^-}} {R \om_c^{\rm{H}^-}} .
\eea

The above result shows that Penning-trap experiments 
comparing the charge-to-mass ratios between  
an antiproton and a proton by using a hydrogen ion as a proxy for the proton
are sensitive not only to the SME coefficients for protons, 
but also provide access to these for electrons. 
From the transformation results presented in 
Appendix~\ref{transformations},
the related coefficients for Lorentz violation in the Sun-centered frame 
are these 81 independent tilde quantities
$\bptw p {J}$,
$\ctw p {(JK)}$,
$\btw p {J(KL)}$,
$\bptws p {J}$,
$\ctws p {(JK)}$,
$\btws p {J(KL)}$,
$\bptw e {J}$,
$\ctw e {(JK)}$,
and
$\btw e {J(KL)}$.
Using expression \rf{ratioH2}
the published results from experiments comparing the charge-to-mass ratios 
between an antiproton and a hydrogen ion
can be adopted to set bounds on the relevant coefficients for Lorentz violation.

\subsection{Experimental sensitivities}
\label{experimental sensitivities}
In this subsection,
we focus on the analysis of two Penning trap experiments comparing the 
charge-to-mass ratios between an antiproton and a proton
and provide the explicit combinations of the tilde coefficients that are sensitive 
to each individual experiment.
Taking the published results we constrain the relevant tilde coefficients
in the Sun-centered frame.

\renewcommand{\arraystretch}{1.5}
\begin{table*}
\caption{
\label{exp-transf}
Combinations of tilde coefficients in the Sun-centered frame 
for both the ATRAP and BASE experiments.}
\setlength{\tabcolsep}{5pt}
\begin{tabular}{llll}
\hline
\hline																						
Experiment	&		Lab. frame		&		Sun-centered frame		&		Harmonic		\\	\hline
ATRAP	&	$	 \bptw {w} {3} 	$	&	$	0.72   \bptw w {Z}	$	&	$	1	$	\\	
	&	$		$	&	$	0.69   \bptw w {X}	$	&	$	\codt	$	\\	
	&	$		$	&	$	0.69   \bptw w {Y} 	$	&	$	\sodt	$	\\	[5pt]
	&	$	 \ctw w {11} +  \ctw w {22}	$	&	$	0.76 (   \ctw w {XX} +   \ctw w {YY} ) + 0.48   \ctw w {ZZ}	$	&	$	1	$	\\	
	&	$		$	&	$	 -1.0   \ctw w {(XZ)}	$	&	$	\codt	$	\\	
	&	$		$	&	$	 -1.0   \ctw w {(YZ)}	$	&	$	\sodt	$	\\	
	&	$		$	&	$	 -0.24 (  \ctw w {XX} -   \ctw w {YY}) 	$	&	$	\ctodt	$	\\	
	&	$		$	&	$	 -0.48   \ctw w {(XY)}	$	&	$	\stodt	$	\\	[5pt]
	&	$	 \btw w {311} +  \btw w {322}	$	&	$	  -0.35 (  \btw w {X(XZ)} +   \btw w {Y(YZ)} -   \btw w {ZZZ}) + 0.55 (  \btw w {ZXX} +   \btw w {ZYY})	$	&	$	1	$	\\	
	&	$		$	&	$	0.44   \btw w {XXX} + 0.61   \btw w {XYY} + 0.33   \btw w {XZZ} - 0.17   \btw w {Y(XY)} - 0.72   \btw w {Z(XZ)} 	$	&	$	\codt	$	\\	
	&	$		$	&	$	 -0.17   \btw w {X(XY)} + 0.61   \btw w {YXX} + 0.44  \btw w {YYY} + 0.33   \btw w {YZZ} - 0.72  \btw w {Z(YZ)}	$	&	$	\sodt	$	\\	 
	&	$		$	&	$	 -0.35 (  \btw w {X(XZ)} -   \btw w {Y(YZ)}) - 0.17 (  \btw w {ZXX} -   \btw w {ZYY})	$	&	$	\ctodt	$	\\	
	&	$		$	&	$	 -0.35 (  \btw w {X(YZ)} +   \btw w {Y(XZ)} +   \btw w {Z(XY)})	$	&	$	\stodt	$	\\	 
	&	$		$	&	$	 -0.08  ( \btw w {XXX} -  \btw w {XYY}) + 0.17   \btw w {Y(XY)}	$	&	$	\cthodt	$	\\	
	&	$		$	&	$	 -0.17    \btw w {X(XY)}  - 0.08 (  \btw w {YXX} -   \btw w {YYY})	$	&	$	\sthodt	$	\\	[10pt]
														
BASE	&	$	 \bptw {w} {3} 	$	&	$	0.35  \bptw w {Z}	$	&	$	1	$	\\	
	&	$		$	&	$	 -0.36  \bptw w {X}  + 0.87  \bptw w {Y}	$	&	$	\codt	$	\\	
	&	$		$	&	$	 -0.87  \bptw w {X}  - 0.36  \bptw w {Y} 	$	&	$	\sodt	$	\\	[5pt]
	&	$	 \ctw w {11} +  \ctw w {22}	$	&	$	0.56 (  \ctw w {XX} +  \ctw w {YY} ) + 0.88  \ctw w {ZZ}	$	&	$	1	$	\\	
	&	$		$	&	$	0.25  \ctw w {(XZ)} - 0.60  \ctw w {(YZ)}	$	&	$	\codt	$	\\	
	&	$		$	&	$	0.60  \ctw w {(XZ)} + 0.25  \ctw w {(YZ)}	$	&	$	\sodt	$	\\	
	&	$		$	&	$	0.31 ( \ctw w {XX} -  \ctw w {YY}) + 0.63  \ctw w {(XY)}	$	&	$	\ctodt	$	\\	
	&	$		$	&	$	 -0.31 ( \ctw w {XX} -  \ctw w {YY}) + 0.62  \ctw w {(XY)}	$	&	$	\stodt	$	\\	[5pt]
	&	$	 \btw w {311} +  \btw w {322}	$	&	$	 -0.30 ( \btw w {X(XZ)} +  \btw w {Y(YZ)} -  \btw w {ZZZ}) + 0.19 ( \btw w {ZXX} +  \btw w {ZYY})	$	&	$	1	$	\\	
	&	$		$	&	$	 -0.12  \btw w {XXX} - 0.38  \btw w {X(XY)} - 0.28  \btw w {XYY} -0.32  \btw w {XZZ} + 0.68  \btw w {YXX} + 0.16  \btw w {Y(XY)} 	$	&	$	\codt	$	\\	
	&	$		$	&	$	\hskip 50pt + 0.29  \btw w {YYY} + 0.76  \btw w {YZZ} + 0.09  \btw w {Z(XZ)} - 0.21  \btw w {Z(YZ)}	$	&	$	 	$	\\	
	&	$		$	&	$	 -0.29  \btw w {XXX} + 0.16  \btw w {X(XY)} -0.68  \btw w {XYY} - 0.76  \btw w {XZZ} - 0.28  \btw w {YXX} + 0.38  \btw w {Y(XY)} 	$	&	$	\sodt	$	\\	 
	&	$		$	&	$	\hskip 50pt - 0.12  \btw w {YYY} - 0.32  \btw w {YZZ} + 0.21  \btw w {Z(XZ)} + 0.09  \btw w {Z(YZ)}	$	&	$	 	$	\\	
	&	$		$	&	$	0.21 ( \btw w {X(XZ)} -  \btw w {Y(YZ)}) + 0.11 ( \btw w {ZXX} -  \btw w {ZYY}) + 0.22 ( \btw w {X(YZ)} +  \btw w {Y(XZ)}  +  \btw w {Z(XY)}) 	$	&	$	\ctodt	$	\\	
	&	$		$	&	$	 -0.22 ( \btw w {X(XZ)} -  \btw w {Y(YZ)}) - 0.11 ( \btw w {ZXX} -  \btw w {ZYY}) + 0.21 ( \btw w {X(YZ)} +  \btw w {Y(XZ)} +  \btw w {Z(XY)}) 	$	&	$	\stodt	$	\\	 
	&	$		$	&	$	 -0.19 ( \btw w {XXX} -  \btw w {XYY}) + 0.38  \btw w {Y(XY)} + 0.16  \btw w {X(XY)}  + 0.08 ( \btw w {YXX} -  \btw w {YYY})	$	&	$	\cthodt	$	\\	
	&	$		$	&	$	 -0.08 ( \btw w {XXX} -  \btw w {XYY}) + 0.16  \btw w {Y(XY)} - 0.38  \btw w {X(XY)}  - 0.19 ( \btw w {YXX} -  \btw w {YYY})	$	&	$	\sthodt	$	\\
\hline
\hline
\end{tabular}
\end{table*}

\subsubsection{The ATRAP experiment}
\label{the atrap experiment}

In the ATRAP experiment located at CERN,
Gabrielse and his collaboration compared the charge-to-mass ratios 
between an antiproton and a proton to a precision of 90 ppt 
using a simultaneously trapped antiproton and hydrogen ion 
in a vertical uniform magnetic field 
$B = 5.85$~T~\cite{ga99}. 
The reported precision was obtained by analyzing the measurements of the 
cyclotron frequencies in a time-averaged way,
so any effects in the difference~\rf{ratioH2} that are dependent of 
sidereal time averaged out. 
This implies that the published precision can be used to constrain only
the tilde coefficients that appear in the constant terms 
in the transformation results listed in Appendix~\ref{transformations}.
In principle, 
a sidereal-variation analysis of the experiment data 
could also be performed to obtain the constraints 
on other components of the tilde coefficients 
that are related to the harmonic terms in the transformation results.  
For the reference of future sidereal-variation analysis of the ATRAP experiment, 
we take $\ch =43.8^\circ$ for the laboratory colatitude
and present in 
Table~\ref{exp-transf} 
the explicit combinations of the tilde coefficients for all the related harmonics
in the transformation results of the tilde quantities
$\bptw w {3}$, 
$\ctw w {11} + \ctw w {22}$,
and
$\btw w {311} + \btw w {322}$.
The table is organized as follows.
The first column specifies the name of the experiment,
and the second column gives the relevant tilde quantities in the laboratory frame.
The corresponding combinations of the tilde coefficients 
in the Sun-centered frame are listed in the third column,
with the associated harmonics displaced in the final column. 


Using expression \rf{ratioH2} together with the reported precision of 90 ppt for 
$(|q|/m)_{\bar{p}}/(|q|/m)_{p} - 1$
and $\om_c^{\rm{H}^-} =  2\pi \times 89.3$ MHz 
for the ATRAP experiment, 
we obtain the following limit in natural units 
\beq
\label{limit-atrap}
| \de \om_c^{\bar{p}} - 1.001 \de \om_c^{p} -2.002  \de \om_c^{e^-}|_{\rm{const}} 
\lsim 3.33 \times 10^{-26}\ \rm{GeV},
\\
\eeq
where the subscript ``const" implies this limit is only for the tilde coefficients appearing 
in the constant terms in the transformations results.

\subsubsection{The BASE experiment}

\renewcommand{\arraystretch}{1.15}
\begin{table}
\caption{
\label{cons}
Constraints on tilde coefficients for Lorentz violation from the ATRAP and the BASE experiments.}
\setlength{\tabcolsep}{5.5pt}
\begin{tabular}{lll}
\hline
\hline																									Coefficient						&				Constraint				&	Experiment	\\	\hline
$	|	 \bptw e {Z}	|				$	&	$	<	1.7	\times 	10^{-17}	$	GeV	&	ATRAP	\\	
$	|	 \ctw e {XX}	|				$	&	$	<	3.2	\times 	10^{-14}	$		&	ATRAP	\\	
$	|	 \ctw e {YY}	|				$	&	$	<	3.2	\times 	10^{-14}	$		&	ATRAP	\\	
$	|	 \ctw e {ZZ}	|				$	&	$	<	2.1	\times 	10^{-14}	$		&	BASE	\\	
$	|	 \btw e {X(XZ)}	|				$	&	$	<	1.2	\times 	10^{-10}	$	 GeV$^{-1}$	&	BASE	\\	
$	|	 \btw e {Y(YZ)}	|				$	&	$	<	1.2	\times 	10^{-10}	$	 GeV$^{-1}$	&	BASE	\\	
$	|	 \btw e {ZZZ}	|				$	&	$	<	1.2	\times 	10^{-10}	$	 GeV$^{-1}$	&	BASE	\\	
$	|	 \btw e {ZXX}	|				$	&	$	<	8.8	\times 	10^{-11}	$	 GeV$^{-1}$	&	ATRAP	\\	
$	|	 \btw e {ZYY}	|				$	&	$	<	8.8	\times 	10^{-11}	$	 GeV$^{-1}$	&	ATRAP	\\	
																			
$	|	 \bptw e {X}	|				$	&	$	<	1.1	\times 	10^{-16}	$	GeV	&	BASE	\\	
$	|	 \bptw e {Y}	|				$	&	$	<	1.1	\times 	10^{-16}	$	GeV	&	"	\\	
$	|	 \ctw e {(XZ)}	|				$	&	$	<	3.0	\times 	10^{-13}	$		&	"	\\	
$	|	 \ctw e {(YZ)}	|				$	&	$	<	3.0	\times 	10^{-13}	$		&	"	\\	
$	|	 \btw e {XXX}	|				$	&	$	<	1.2	\times 	10^{-9}	$	 GeV$^{-1}$	&	"	\\	
$	|	 \btw e {X(XY)}	|				$	&	$	<	9.3	\times 	10^{-10}	$	 GeV$^{-1}$	&	"	\\	
$	|	 \btw e {XYY}	|				$	&	$	<	5.2	\times 	10^{-10}	$	 GeV$^{-1}$	&	"	\\	
$	|	 \btw e {XZZ}	|				$	&	$	<	4.6	\times 	10^{-10}	$	 GeV$^{-1}$	&	"	\\	
$	|	 \btw e {YXX}	|				$	&	$	<	5.2	\times 	10^{-10}	$	 GeV$^{-1}$	&	"	\\	
$	|	 \btw e {Y(XY)}	|				$	&	$	<	9.3	\times 	10^{-10}	$	 GeV$^{-1}$	&	"	\\	
$	|	 \btw e {YYY}	|				$	&	$	<	1.2	\times 	10^{-9}	$	 GeV$^{-1}$	&	"	\\	
$	|	 \btw e {YZZ}	|				$	&	$	<	4.6	\times 	10^{-10}	$	 GeV$^{-1}$	&	"	\\	
$	|	 \btw e {Z(XZ)}	|				$	&	$	<	1.7	\times 	10^{-9}	$	 GeV$^{-1}$	&	"	\\	
$	|	 \btw e {Z(YZ)}	|				$	&	$	<	1.7	\times 	10^{-9}	$	 GeV$^{-1}$	&	"	\\	[5pt]
																	 		
$	|	 \bptw p {Z}	|,	|	 \bptws p {Z}	|	$	&	$	<	1.2	\times 	10^{-10}	$	GeV	&	ATRAP	\\	
$	|	 \ctw p {XX}	|,	|	 \ctws p {XX}	|	$	&	$	<	1.2	\times 	10^{-10}	$		&	ATRAP	\\	
$	|	 \ctw p {YY}	|,	|	 \ctws p {YY}	|	$	&	$	<	1.2	\times 	10^{-10}	$		&	ATRAP	\\	
$	|	 \ctw p {ZZ}	|,	|	 \ctws p {ZZ}	|	$	&	$	<	7.9	\times 	10^{-11}	$		&	BASE	\\	
$	|	 \btw p {X(XZ)}	|,	|	 \btws p {X(XZ)}	|	$	&	$	<	2.4	\times 	10^{-10}	$	 GeV$^{-1}$	&	BASE	\\	
$	|	 \btw p {Y(YZ)}	|,	|	 \btws p {Y(YZ)}	|	$	&	$	<	2.4	\times 	10^{-10}	$	 GeV$^{-1}$	&	BASE	\\	
$	|	 \btw p {ZZZ}	|,	|	 \btws p {ZZZ}	|	$	&	$	<	2.4	\times 	10^{-10}	$	 GeV$^{-1}$	&	BASE	\\	
$	|	 \btw p {ZXX}	|,	|	 \btws p {ZXX}	|	$	&	$	<	1.8	\times 	10^{-10}	$	 GeV$^{-1}$	&	ATRAP	\\	
$	|	 \btw p {ZYY}	|,	|	 \btws p {ZYY}	|	$	&	$	<	1.8	\times 	10^{-10}	$	 GeV$^{-1}$	&	ATRAP	\\	
																			
$	|	 \bptw p {X}	|,	|	 \bptws p {X}	|	$	&	$	<	7.2	\times 	10^{-10}	$	GeV	&	BASE	\\	
$	|	 \bptw p {Y}	|,	|	 \bptws p {Y}	|	$	&	$	<	7.2	\times 	10^{-10}	$	GeV	&	"	\\	
$	|	 \ctw p {(XZ)}	|,	|	 \ctws p {(XZ)}	|	$	&	$	<	1.1	\times 	10^{-9}	$		&	"	\\	
$	|	 \ctw p {(YZ)}	|,	|	 \ctws p {(YZ)}	|	$	&	$	<	1.1	\times 	10^{-9}	$		&	"	\\	
$	|	 \btw p {XXX}	|,	|	 \btws p {XXX}	|	$	&	$	<	2.4	\times 	10^{-9}	$	 GeV$^{-1}$	&	"	\\	
$	|	 \btw p {X(XY)}	|,	|	 \btws p {X(XY)}	|	$	&	$	<	1.9	\times 	10^{-9}	$	 GeV$^{-1}$	&	"	\\	
$	|	 \btw p {XYY}	|,	|	 \btws p {XYY}	|	$	&	$	<	1.1	\times 	10^{-9}	$	 GeV$^{-1}$	&	"	\\	
$	|	 \btw p {XZZ}	|,	|	 \btws p {XZZ}	|	$	&	$	<	9.3	\times 	10^{-9}	$	 GeV$^{-1}$	&	"	\\	
$	|	 \btw p {YXX}	|,	|	 \btws p {YXX}	|	$	&	$	<	1.1	\times 	10^{-9}	$	 GeV$^{-1}$	&	"	\\	
$	|	 \btw p {Y(XY)}	|,	|	 \btws p {Y(XY)}	|	$	&	$	<	1.9	\times 	10^{-9}	$	 GeV$^{-1}$	&	"	\\	
$	|	 \btw p {YYY}	|,	|	 \btws p {YYY}	|	$	&	$	<	2.4	\times 	10^{-9}	$	 GeV$^{-1}$	&	"	\\	
$	|	 \btw p {YZZ}	|,	|	 \btws p {YZZ}	|	$	&	$	<	9.3	\times 	10^{-10}	$	 GeV$^{-1}$	&	"	\\	
$	|	 \btw p {Z(XZ)}	|,	|	 \btws p {Z(XZ)}	|	$	&	$	<	3.4	\times 	10^{-9}	$	 GeV$^{-1}$	&	"	\\	
$	|	 \btw p {Z(YZ)}	|,	|	 \btws p {Z(YZ)}	|	$	&	$	<	3.4	\times 	10^{-9}	$	 GeV$^{-1}$	&	"	\\												
\hline
\hline
\end{tabular}
\end{table}

Another Penning-trap experiment at CERN
by the BASE collaboration 
recently improved the same comparison 
to the record sensitivity of 69 
ppt~\cite{ul15}.
Since the BASE experiment also used a hydrogen ion as a proxy of a proton,
the expression \rf{ratioH2} still holds.
The trap used a horizontal magnetic field $B = 1.946$ T oriented 
$60^\circ$ east of north.
This implies that both matrices \rf{rot} and \rf{euler} are needed for 
determining the combination of the tilde coefficients in the Sun-centered frame.
The corresponding Euler angles for a horizontal magnetic field 
with an angle~$\th$ from the local south in the counterclockwise direction
are found to be $(\al, \be, \ga) =  (\th, \pi/2, 0)$.
Taking $\th=2 \pi/3$ and $\ch =43.8^\circ$ for the BASE experiment
we also include in
Table~\ref{exp-transf} 
the related explicit transformation results 
for the tilde quantities
$\bptw w {3}$, 
$\ctw w {11} + \ctw w {22}$,
and
$\btw w {311} + \btw w {322}$.

Different from the ATRAP experiment, 
the experimental data of the charge-to-mass ratio comparison 
for the BASE experiment 
were analyzed to search for both time-averaged effects
and sidereal variations in the first harmonic of the Earth's rotation frequency,
so the reported sensitivities from the experiment 
can be taken to set bounds on both the tilde coefficients appearing in the constant terms 
and these in the first harmonic of the oscillations.  
Using 69 ppt for the time-averaged precision
and 720 ppt for the limit of the first harmonic amplitude, 
together with $\om_c^{\rm{H}^-} =  2\pi \times 29.635$ MHz from the experiment,
expression~\rf{ratioH2} yields 
\beq
\label{limit-base}
| \de \om_c^{\bar{p}} - 1.001 \de \om_c^{p} -2.002  \de \om_c^{e^-}|_{\rm{const}} 
\lsim 8.46 \times 10^{-27}\ \rm{GeV}
\\
\eeq
and 
\beq
\label{limit-base-1st}
| \de \om_c^{\bar{p}} - 1.001 \de \om_c^{p} -2.002  \de \om_c^{e^-}|_{\rm{1st}} 
\lsim 8.83 \times 10^{-26}\ \rm{GeV}
\\
\eeq
in natural units,
where the subscript ``const" in limit \rf{limit-base} 
has the same meaning as what in \rf{limit-atrap},
while the subscript ``1st" in limit \rf{limit-base-1st} represents
the amplitude of the first harmonic in the sidereal variation.   

Some intuition about the scope of the constraints 
on the individual components of the tilde coefficients
appearing in limits 
\rf{limit-atrap}, \rf{limit-base}, and \rf{limit-base-1st}
can be obtained by 
assuming that only one individual tilde coefficient to be nonzero 
at a time and extracting its resulting constraint,
which is a common practice adopted in many subfields searching for 
Lorentz and CPT 
violation~\cite{tables}.
Using the limit~\rf{limit-atrap} from the ATRAP experiment, 
a total of 27 independent tilde coefficients for Lorentz violation
are constrained.
For the BASE experiment,
69 constraints are obtained on the independent tilde 
coefficients for Lorentz violation 
from limits \rf{limit-base} and \rf{limit-base-1st}.
To summarize the results,
we list in 
\tab{cons}
the individual constraints on the tilde coefficients in the Sun-centered frame,
with the first column listing the individual components, 
the second column presenting the corresponding constraint 
on the modulus of each one,
and the third column specifying the related experiment. 
Note that when a component of the tilde coefficients is constrained 
by both the ATRAP and the BASE experiment,
we only include the more stringent one in 
\tab{cons}.


\tab{cons} shows that
69 of the 81 independent components 
of the tilde coefficients
that are related to Penning-trap experiments 
comparing charge-to-mass ratios of an antiproton and a proton
can be bounded using the published results.
The other 12 components lie in the second and third harmonics 
of the transformation results given by 
Appendix~\ref{transformations}
and
Table~\ref{exp-transf}.
Therefore, 
in order to extract all the related constraints on the tilde coefficients,
a full sidereal-variation analysis must be performed on the experimental data. 
The above 69 tilde coefficients listed in 
\tab{cons} 
are analyzed for the first time 
as they lie in different coefficient space compared to existing ones 
that are related to Penning-trap experiments comparing the $g$ factors 
between particles and 
antiparticles~\cite{16dk}.

\section{Summary}
\label{summary}

In this work,
we applied the general theory of quantum electrodynamics
with Lorentz- and CPT-violating operators of mass dimensions up to six
to Penning-trap experiments comparing the charge-to-mass ratios 
between antiprotons and protons. 
Using perturbation theory,
we derived the dominant Lorentz- and CPT-violating contributions 
\rf{linear-en} and \rf{linear-ens}
to the energy levels 
of the confined particles and antiparticles,
which enabled us to determine the corresponding cyclotron frequency shifts
 \rf{wcshift} and \rf{wcsshift}.
Relating the experimental interpreted charge-to-mass ratio comparisons
to the cyclotron frequency shifts, 
we addressed the issue of a CPT test
and concluded that it depends on the context of the relevant theory.
We found in Eq. \rf{ratioH2} that the coefficients for Lorentz violation
that are sensitive to the charge-to-mass ratio comparison
between antiprotons and protons
are the 81 independent tilde quantities 
$\bptw p {J}$,
$\ctw p {(JK)}$,
$\btw p {J(KL)}$,
$\bptws p {J}$,
$\ctws p {(JK)}$,
$\btws p {J(KL)}$,
$\bptw e {J}$,
$\ctw e {(JK)}$,
and
$\btw e {J(KL)}$
in the Sun-centered frame. 
Using published results from the ATRAP
and the BASE experiments,
we obtained first-time constraints on 69 of them
and summarized the results in 
\tab{cons}.
To set bounds to the other 12 components of the tilde coefficients for Lorentz violation,
a full sidereal-variation analysis of the experimental data is required. 
The high-precision measurements and excellent coverage of the SME coefficients
offered by current and forthcoming Penning-trap experiments provide
strong motivations to continue the searches for possible 
Lorentz- and CPT-violating signals.

\section{Acknowledgements}

We thank V.A.\ Kosteleck\'y for useful discussions
and K.\ Andresen for reading the manuscript. 
This work~was supported in part by the
Gettysburg College Cross-Disciplinary Science Institute (X-SIG).

\appendix
 
\onecolumngrid
\section{Full Lagrange density for $d\leq 6$}
\label{lag}

The full Lagrange density \rf{fermlag}
with Lorentz-violating operators of mass dimensions $d\leq 6$
can be taken as the sum of the conventional Lorentz-invariant
QED Lagrange density $\cl_0$ and a series of Lorentz-violating terms $\cl^{(d)}$ 
of mass dimension $d$,
$
\cl_\ps =
\cl_0 
+ \cl^{(3)} 
+ \cl^{(4)} 
+ \cl^{(5)} 
+ \cl^{(6)} 
+ \ldots .
\label{nrlag}
$
The explicit results were given in 
Ref.~\cite{16dk}.
Here, 
we reproduce these terms. 

The pieces in the minimal SME Lagrange density $\cl^{(3)}$ and $\cl^{(4)}$ are 
\bea
\cl^{(3)} &=&
- \acm 3 \mu \psb \ga_\mu \ps
- \bcm 3 \mu \psb \ga_5 \ga_\mu \ps
- \half \Hcm 3 \mn \psb \si_\mn \ps, 
\\
\cl^{(4)} &=&
\half \ccm 4 {\ma} \psb \ga_\mu iD_\al \ps
+ \half \dcm 4 {\ma} \psb \ga_5 \ga_\mu iD_\al \ps
+ \half \ecm 4 {\al} \psb iD_\al \ps
+ \half i \fcm 4 {\al} \psb \ga_5 iD_\al \ps
+ \quar \gcm 4 {\mna} \psb \si_\mn iD_\al \ps
+ {\rm H.c.}
\label{lag4}
\eea

The dimension-five Lagrange density $\cl^{(5)}$ can be classified into two kinds,
$
\cl^{(5)} = \cl^{(5)}_D +\cl^{(5)}_F ,
$
with $\cl^{(5)}_D$ containing symmetrized covariant derivatives $D_\al$ 
and $\cl^{(5)}_F$ involving the antisymmetric electromagnetic field strength $F_\ab$,
given by
\bea
\cl^{(5)}_D &=&
- \half \mc 5 \ab \psb iD_{(\al} iD_{\be)} \ps
- \half i \mfc 5 \ab \psb \ga_5 iD_{(\al} iD_{\be)} \ps
- \half \ac 5 \mab \psb \ga_\mu iD_{(\al} iD_{\be)} \ps
- \half \bc 5 \mab \psb \ga_5 \ga_\mu iD_{(\al} iD_{\be)} \ps
\nn\\
&&
- \quar \Hc 5 \mnab \psb \si_\mn iD_{(\al} iD_{\be)} \ps
+ {\rm H.c.} ,
\\
\cl^{(5)}_F &=&
- \half \mcf 5 \ab F_\ab \psb \ps
- \half i \mfcf 5 \ab F_\ab \psb \ga_5 \ps
- \half \acf 5 \mab F_\ab \psb \ga_\mu \ps
- \half \bcf 5 \mab F_\ab \psb \ga_5 \ga_\mu \ps
- \quar \Hcf 5 \mnab F_\ab \psb \si_\mn \ps.
\nn \\
\label{l5f}
\eea

For $d=6$,
there are three types of terms, 
$
\cl^{(6)} = \cl^{(6)}_D + \cl^{(6)}_F + \cl^{(6)}_{\prt F} ,
$
where
\bea
\cl^{(6)}_D &=&
\half \cc 6 {\mabc} 
\psb \ga_\mu iD_{(\al} iD_\be iD_{\ga)} \ps 
+ \half \dc 6 {\mabc} 
\psb \ga_5 \ga_\mu iD_{(\al} iD_\be iD_{\ga)} \ps 
+ \half \ec 6 {\abc} 
\psb iD_{(\al} iD_\be iD_{\ga)} \ps 
\nn\\
&&
+ \half i \fc 6 {\abc} 
\psb \ga_5 iD_{(\al} iD_\be iD_{\ga)} \ps 
+ \quar \gc 6 {\mnabc} 
\psb \si_\mn iD_{(\al} iD_\be iD_{\ga)} \ps 
+ {\rm H.c.} ,
\\
\cl^{(6)}_F &=&
\quar \ccf 6 {\mabc} F_{\bec} 
\big(
\psb \ga_\mu iD_{\al} \ps 
+ {\rm H.c.}
\big)
+ \quar \dcf 6 {\mabc} F_{\bec} 
\big(
\psb \ga_5 \ga_\mu iD_{\al} \ps 
+ {\rm H.c.}
\big)
+ \quar \ecf 6 {\abc} F_{\bec} 
\big(
\psb iD_{\al} \ps 
+ {\rm H.c.}
\big)
\nn\\
&&
+ \quar i \fcf 6 {\abc} F_{\bec} 
\big(
\psb \ga_5 iD_{\al} \ps 
+ {\rm H.c.}
\big)
+ \frac 1 8 \gcf 6 {\mnabc} F_{\bec} 
\big(
\psb \si_\mn iD_{\al} \ps 
+ {\rm H.c.} 
\big),
\\
\cl^{(6)}_{\prt F} &=&
- \half \mcpf 6 \abc \prt_\al F_\bec ~\psb \ps
- \half i \mfcpf 6 \abc \prt_\al F_\bec ~\psb \ga_5 \ps
- \half \acpf 6 \mabc \prt_\al F_\bec ~\psb \ga_\mu \ps
- \half \bcpf 6 \mabc \prt_\al F_\bec ~\psb \ga_5 \ga_\mu \ps
\nn\\
&&
- \quar \Hcpf 6 \mnabc \prt_\al F_\bec ~\psb \si_\mn \ps.
\label{l6df}
\eea

In the above expressions,
dimension superscripts for the minimal-SME coefficients are omitted. 
Coefficients with subscript $F$ or $\partial F$ are contracted with operators involving
the electromagnetic field strength or its derivative,
where indices $\mu, \nu$ are associated with spin properties,
while $\al, \be, \ga$ are related to covariant momenta including field strengths. 
Parentheses on $n$ indices represent symmetrization with a factor of $1/n!$.  
The properties of the coefficients for Lorentz violation appearing above 
are listed in Table I in Ref.~\cite{16dk}.

\section{Perturbative energy shifts}
\label{perturbative energy shifts}

The perturbative energy shifts $\de E_{n, \pm 1}^w$ due to Lorentz and CPT violation 
for a fermion species $w$ of mass $m_w$ and charge $q=\si |q|$ 
in a magnetic field 
$\mbf B = B \hat x_3$ in the apparatus frame 
can be obtained by applying perturbation calculations using Eq.~\rf{deE}.
The analysis is performed with Lorentz- and CPT-violating operators appearing in
$\cl^{(3)}$, $\cl^{(4)}$, $\cl^{(5)}_D$, and $\cl^{(6)}_D$,
listed above in 
Appendix \ref{lag}.
Following the discussion in subsection \ref{energy shifts}, 
we find

\bea
\label{fullen}
\de E_{n, \pm 1}^w
&=&
a_w^{0} 
\mp \si b_w^{3} \dfrac{m_w}{E_{n, \pm 1}^w }
\mp \si H_w^{12}
- c_w^{00} E_{n, \pm 1}^w - (c_w^{11} + c_w^{22}) \dfrac{(2n+1 \mp \si )}{2 E_{n, \pm 1}^w} |qB|
\pm \si d_w^{30}m_w-e_w^{0}m_w
\nn \\
&&
\mp \si (g_w^{012} - g_w^{021}) \dfrac{(2n+1 \mp \si )}{2E_{n, \pm 1}^w} |qB| \pm \si g_w^{120} E_{n, \pm 1}^w
+ a_w^{(5)000} (E_{n, \pm 1}^w)^2 
\nn \\
&& 
+ (a_w^{(5)011} + a_w^{(5)022}) (\dfrac{2n+1 \mp \si }{2} \pm \si \dfrac{m_w}{2E_{n, \pm 1}^w} ) |qB| 
+ (a_w^{(5)101} + a_w^{(5)202}) (2n+1 \mp \si ) |qB|
\mp \si b_w^{(5)300}m_wE_{n, \pm 1}^w
\nn \\
&&
\mp \si (b_w^{(5)311} + b_w^{(5)322}) ( \dfrac{2n+1 \mp \si }{2} \dfrac{m_w}{E_{n, \pm 1}^w} \pm \dfrac{1}{2} \si  ) |qB|
\pm \si (H_w^{(5)0102} - H_w^{(5)0201}) (2n+1\mp \si) |qB| 
\nn \\
&&
\mp \si H_w^{(5)1200} (E_{n, \pm 1}^w)^2
\mp \si (H_w^{(5)1211} + H_w^{(5)1222}) (\dfrac{2n+1 \mp \si }{2} \pm \si \dfrac{m_w}{2 E_{n, \pm 1}^w} ) |qB|
+ m_w^{(5)00} m_w E_{n, \pm 1}^w
\nn \\
&&
+ (m_w^{(5)11} + m_w^{(5)22}) ( \dfrac{2n+1 \mp \si}{2} \dfrac{m_w}{E_{n, \pm 1}^w} \pm \dfrac{1}{2}  \si) |qB|
-c_w^{(6)0000} (E_{n, \pm 1}^w)^3
\nn \\
&&
- 3(c_w^{(6)0011} + c_w^{(6)0022}) (\dfrac{2n+1 \mp \si}{2} E_{n, \pm 1}^w \pm  \dfrac{1}{2} \si m_w ) |qB|
- 3(c_w^{(6)1001} + c_w^{(6)2002})\dfrac{2n+1 \mp \si }{2} E_{n, \pm 1}^w |qB|
\nn \\
&&
- 3(c_w^{(6)1111} + c_w^{(6)2222} + c_w^{(6)1122}+c_w^{(6)2112}) \dfrac{(2n+1 \mp \si )^2}{8E_{n, \pm 1}^w} |qB|^2
\pm \si d_w^{(6)3000}m_w (E_{n, \pm 1}^w)^2
\nn \\
&&
+ 3(d_w^{(6)3011} + d_w^{(6)3022})  ( \dfrac{2n+1 \mp \si }{2} m_w \pm \dfrac{1}{2} \si E_{n, \pm 1}^w ) |qB|
- e_w^{(6)000} m_w (E_{n, \pm 1}^w)^2
\nn \\
&&
- 3(e_w^{(6)011} + e_w^{(6)022}) ( \dfrac{2n+1 \mp \si }{2} m_w \pm \dfrac{1}{2} \si E_{n, \pm 1}^w  ) |qB|
\mp 3 \si (g_w^{(6)01002} - g_w^{(6)02001}) \dfrac{2n+1\mp \si}{2} E_{n, \pm 1}^w |qB|
\nn \\
&&
\mp 3 \si (g_w^{(6)01112} + g_w^{(6)01222} - g_w^{(6)02122} - g_w^{(6)02111}) \dfrac{(2n+1 \mp \si )^2}{8E_{n, \pm 1}^w} |qB|^2
\pm \si g_w^{(6)12000} (E_{n, \pm 1}^w)^3 
\nn \\
&&
\pm 3 \si (g_w^{(6)12011} + g_w^{(6)12022}) ( \dfrac{2n+1 \mp \si }{2} E_{n, \pm 1}^w \pm \dfrac{1}{2} \si m_w ) |qB| ,
\eea
where signs $\pm$ denote the spin-up and spin-down states,
respectively,
and the unperturbed positive eigenenergies are given by
$E_{n, \pm 1}^w = \sqrt{m_w^2 + (2n+1 \pm 1)|qB|}$.
The additional energy shift contributions from $\cl^{(5)}_F$ and $\cl^{(6)}_F$
can be obtained by the substitutions~(40) in
Ref.~\cite{16dk},
while $\cl^{(6)}_{\prt F}$ has no energy shift contributions 
as $\prt_\al F_{\be \ga}=0$ for a uniform magnetic field in a Penning trap.

\section{Transformations}
\label{transformations}

In this Appendix,
we present the explicit relations between the coefficients for Lorentz violation 
$\bptw w {3}$, 
$\ctw w {11} + \ctw w {22}$,
$\btw w {311} + \btw w {322}$,
$\bptws w {3}$, 
$\ctws w {11} + \ctws w {22}$,
and
$\btws w {311} + \btws w {322}$
in the apparatus frame 
and the constant ones in the Sun-centered frame.

For a Penning-trap experiment with a vertical upward magnetic field, 
applying transformation \rf{axis} with $R^{aj}$ being the identity matrix gives
\bea
\bptw w {3} &=& 
\codt  
\bptw w {X} \sin\ch 
+ \sodt
\bptw w {Y} \sin\ch 	
+ 
\bptw w {Z} \cos\ch ,
\eea

\bea
\ctw w {11} + \ctw w {22} &=& 
\ctodt  
\Big(  - \half (\ctw w {XX} -  \ctw w {YY}) \sin^2 \ch \Big)
+ \stodt
\left(  -  \ctw w {(XY)} \sin^2 \ch	 \right)
\nn \\
&& 
+  \codt  
\left(  - \ctw w {(XZ)}  \sin 2\ch \right)
+ \sodt
\left(  - \ctw w {(YZ)}  \sin 2\ch  \right)
\nn \\
&&
+ 
\frac{1}{4} (\ctw w {XX} + \ctw w {YY})(3+\cos 2\ch) + \ctw w {ZZ} \sin^2\ch  ,
\eea

\bea
\btw w {311} + \btw w {322} &=& 
\cthodt  
\left( [-\frac{1}{4} (\btw w {XXX} - \btw w {XYY}) + \half \btw w {Y(XY)}  ] \sin^3 \ch \right)
\nn \\
&&
+ \sthodt
\left( [- \half \btw w {X(XY)}  - \frac{1}{4} (\btw w {YXX} - \btw w {YYY}) ] \sin^3 \ch	\right)
\nn \\
&&
+\ctodt  
\left(  [ - \btw w {X(XZ)} + \btw w {Y(YZ)} - \half (\btw w {ZXX} - \btw w {ZYY}) ] \cos\ch \sin^2 \ch \right)
\nn \\
&&
+ \stodt
\left( [- \btw w {X(YZ)} - \btw w {Y(XZ)} - \btw w {Z(XY)} ] \cos\ch \sin^2 \ch \right)
\nn \\
&& 
+  \codt  
\Big( \frac{1}{8} \btw w {XXX} (5+3 \cos 2\ch)\sin\ch + \frac{1}{8} \btw w {XYY} (7+ \cos 2\ch)\sin\ch + \btw w {XZZ} \sin^3\ch 
\nn \\
&&
\hskip 60pt
- \frac{1}{2} \btw w {Y(XY)} \sin^3\ch - 2 \btw w {Z(XZ)} \cos^2\ch \sin\ch \Big)
\nn \\
&&
+ \sodt
\Big( - \frac{1}{2} \btw w {X(XY)} \sin^3\ch +  \frac{1}{8} \btw w {YXX} (7+ \cos 2\ch)\sin\ch +  \frac{1}{8} \btw w {YYY} (5+ 3\cos 2\ch)\sin\ch 
\nn \\
&&
\hskip 60pt
+  \btw w {YZZ} \sin^3\ch -  2\btw w {Z(YZ)} \cos^2\ch \sin\ch \Big)
\nn \\
&&
- (\btw w {X(XZ)} + \btw w {Y(YZ)} - \btw w {ZZZ} )\cos\ch \sin^2\ch
+ (\btw w {ZXX} + \btw w {ZYY}) \cos\ch \cos2\ch  .
\eea

In the case where the trap uses a horizontal magnetic field
with an angle $\th$ from the local south in the counterclockwise direction,
the corresponding Euler angles relating the apparatus frame to
the standard laboratory frame discussed in 
subsection~\ref{transformation under rotations} 
are found to be 
$(\al, \be, \ga) =  (\th, \pi/2, 0)$.
Substituting this to the matrix \rf{euler} and applying the transformation~\rf{axis}
give the following relations,
\bea
\bptw w {3} &=& 
\codt  
\left( \bptw w {X} \cos\th \cos\ch + \bptw w {Y} \sin\th \right)
+ \sodt
\left( - \bptw w {X} \sin\th + \bptw w {Y} \cos\th \cos\ch \right)
- \bptw w {Z} \cos\th \sin\ch , 
\eea

\bea
\ctw w {11} + \ctw w {22} &=& 
\ctodt  
\left(  \frac{1}{8} (\ctw w {XX} -  \ctw w {YY}) (1-3\cos2\th - 2\cos^2\th \cos2\ch) - \ctw w {(XY)} \cos\ch \sin 2\th \right)
\nn \\
&& 
+ \stodt
\left(   \half (\ctw w {XX} -  \ctw w {YY}) \cos\ch \sin 2\th + \frac{1}{4} \ctw w {(XY)} (1-3\cos2\th - 2\cos^2\th \cos2\ch)	 \right)
\nn \\
&& 
+  \codt  
\left( \ctw w {(XZ)}  \cos^2\th \sin2\ch + \ctw w {(YZ)} \sin2\th \sin\ch \right)
\nn \\
&& 
+ \sodt
\left(  - \ctw w {(XZ)} \sin2\th \sin\ch + \ctw w {(YZ)}  \cos^2\th \sin2\ch  \right)
\nn \\
&&
+ 
\half(\ctw w {XX} + \ctw w {YY}) (\cos^2\th + \cos^2\ch \sin^2\th + \sin^2\ch ) + \ctw w {ZZ} (\cos^2\ch + \sin^2\th \sin^2\ch) ,
\eea

\bea
&&
\btw w {311} + \btw w {322}
\nn \\
 &=& 
\cthodt  
\Big( [\frac{1}{64} (\btw w {XXX} -\btw w {XYY}) - \frac{1}{32} \btw w {Y(XY)}] [3 (\cos \theta -5 \cos 3 \theta ) \cos \chi -4 \cos ^3\theta  \cos 3 \chi ]
\nn \\
&&
\hskip 55pt
+ [\frac{1}{16}\btw w {X(XY)} + \frac{1}{32} (\btw w {YXX} -\btw w {YYY})] [3\sin \theta (1-4 \cos ^2 \theta  \cos 2 \chi) - 5 \sin 3 \theta ]  \Big)
\nn \\
&&
+ \sthodt
\Big( [- \frac{1}{32} (\btw w {XXX} - \btw w {XYY}) + \frac{1}{16} \btw w {Y(XY)}] [3\sin \theta (1-4 \cos ^2 \theta  \cos 2 \chi) - 5 \sin 3 \theta ]
\nn \\
&&
\hskip 60pt
+ [\frac{1}{32} \btw w {X(XY)} + \frac{1}{64} (\btw w {YXX} -\btw w {YYY})]  [3 (\cos \theta -5 \cos 3 \theta ) \cos \chi -4 \cos ^3\theta  \cos 3 \chi ] \Big)
\nn \\
&&
+\ctodt  
\Big( [-\frac{1}{16} \btw w {X(XZ)} + \frac{1}{16}  \btw w {Y(YZ)} - \frac{1}{32} (\btw w {ZXX} -\btw w {ZYY})] [(\cos \theta -5 \cos 3 \theta ) \sin \chi -4 \cos ^3\theta  \sin 3 \chi ]
\nn \\
&&
\hskip 60pt
+(\btw w {X(YZ)} +\btw w {Y(XZ)} +\btw w {Z(XY)} )\sin \theta\cos^2\theta\sin 2\chi \Big)  
\nn \\
&&
+ \stodt
\Big( [-\btw w {X(XZ)} +\btw w {Y(YZ)} - \half (\btw w {ZXX} - \btw w {ZYY})] \sin \theta\cos^2\theta\sin 2\chi
\nn \\
&&
\hskip 60pt
- (\frac{1}{16} \btw w {X(YZ)} + \frac{1}{16} \btw w {Y(XZ)} + \frac{1}{16} \btw w {Z(XY)} ) [(\cos \theta -5 \cos 3 \theta ) \sin \chi -4 \cos ^3\theta  \sin 3 \chi ] \Big)
\nn \\
&& 
+  \codt  
\nn \\
&&
\hskip 20pt 
\times
\Big( \frac{1}{16} \btw w {XXX} \cos \theta \cos \chi (-6 \cos ^2 \theta \cos 2 \chi +3 \cos 2 \theta +7)
- \frac{1}{2} \btw w {X(XY)}  \sin \theta (\cos ^2 \theta  \cos 2 \chi +\sin ^2 \theta  \cos ^2\chi +\sin ^2\chi )
\nn \\
&&
\hskip 35pt 
+ \frac{1}{16} \btw w {XYY} \cos \theta\cos \chi(-2 \cos^2\theta\cos 2\chi +\cos 2\theta+13)
+\btw w {XZZ} \cos \theta\cos \chi(\sin^2\theta\sin^2\chi+\cos^2\chi)
\nn \\
&&
\hskip 35pt
+ \frac{1}{32} \btw w {YXX} [\sin \theta(25-4 \cos^2\theta\cos 2\chi )+\sin 3 \theta]
+ \frac{1}{16} \btw w {Y(XY)} \cos \theta[(\cos 2\theta-7) \cos \chi-2 \cos^2\theta\cos 3 \chi]
\nn \\
&&
\hskip 35pt
+ \frac{1}{32} \btw w {YYY} [\sin \theta(11-12 \cos^2\theta\cos 2\chi )+3 \sin 3 \theta]
+\btw w {YZZ} \sin \theta(\sin^2\theta\sin^2\chi+\cos^2\chi)
\nn \\
&&
\hskip 35pt
-2\btw w {Z(XZ)} \cos^3\theta\sin^2\chi\cos \chi
-2\btw w {Z(YZ)} \sin \theta\cos^2\theta\sin^2\chi \Big)
\nn \\
&&
+ \sodt
\nn \\
&&
\hskip 20pt 
\times
\Big(\frac{1}{32} \btw w {XXX} [\sin \theta(12 \cos^2\theta\cos 2\chi -11)-3 \sin 3 \theta]
+ \frac{1}{16} \btw w {X(XY)} \cos \theta[(\cos 2\theta-7) \cos \chi-2 \cos^2\theta\cos 3 \chi]
\nn \\
&&
\hskip 35pt
+ \frac{1}{32} \btw w {XYY} [\sin \theta(4 \cos^2\theta\cos 2\chi -25)-\sin 3 \theta]
-\btw w {XZZ} \sin \theta(\sin^2\theta\sin^2\chi+\cos^2\chi)
\nn \\
&&
\hskip 35pt
+\frac{1}{16} \btw w {YXX} \cos \theta\cos \chi(-2 \cos^2\theta\cos 2\chi +\cos 2\theta+13)
+ \frac{1}{2} \btw w {Y(XY)} \sin \theta (\cos^2\theta\cos 2\chi +\sin^2\theta\cos^2\chi+\sin^2\chi)
\nn \\
&&
\hskip 35pt
+ \frac{1}{16} \btw w {YYY} \cos \theta\cos \chi(-6 \cos^2\theta\cos 2\chi +3 \cos 2\theta+7)
+\btw w {YZZ} \cos \theta\cos \chi(\sin^2\theta\sin^2\chi+\cos^2\chi)
\nn \\
&&
\hskip 35pt
+2\btw w {Z(XZ)} \sin \theta\cos^2\theta\sin^2\chi
-2\btw w {Z(YZ)} \cos^3\theta\sin^2\chi\cos \chi \Big)
\nn \\
&&
+ 
\frac{1}{8} (\btw w {X(XZ)} + \btw w {Y(YZ)} - \btw w {ZZZ}) \cos \theta (-4 \cos 2\theta \sin^3 \chi + 5 \sin \chi + \sin 3\chi)
\nn \\
&&
+ \frac{1}{16} (\btw w {ZXX} + \btw w {ZYY}) [2 \cos^3 \theta \sin 3\chi - \cos \theta \sin \chi (3 \cos 2\theta + 11 )].
\eea

The corresponding transformation results for the starred tilde quantities 
$\bptws w {3}$, 
$\ctws w {11} + \ctws w {22}$,
and
$\btws w {311} + \btws w {322}$
have the same form as these given above. 

\twocolumngrid


\begin{thebibliography}{99}

\bibitem{ksp}
V.A.\ Kosteleck\'y and S.\ Samuel,
Phys.\ Rev.\ D {\bf 39}, 683 (1989);
V.A.\ Kosteleck\'y and R.\ Potting,
Nucl.\ Phys.\ B {\bf 359}, 545 (1991);
Phys.\ Rev.\ D {\bf 51}, 3923 (1995). 

\bibitem{ck}
D.\ Colladay and V.A.\ Kosteleck\'y,
Phys.\ Rev.\ D {\bf 55}, 6760 (1997);
Phys.\ Rev.\ D {\bf 58}, 116002 (1998).

\bibitem{owg}
O.W.\ Greenberg,
Phys.\ Rev.\ Lett.\ {\bf 89}, 231602 (2002).

\bibitem{tables}
{\it Data Tables for Lorentz and CPT Violation,}
V.A.\ Kosteleck\'y and N.\ Russell,
Rev.\ Mod.\ Phys.\ {\bf 83}, 11 (2011);
2020 edition arXiv:0801.0287v13.

\bibitem{ul15}
S.\ Ulmer \etal, 
Nature {\bf 524}, 196 (2015).

\bibitem{vd87}
R.S.\ Van Dyck, Jr., P.B.\ Schwinberg, and H.G.\ Dehmelt, 
Phys.\ Rev.\ Lett.\ {\bf 59}, 26 (1987);
Phys.\ Rev.\ D {\bf 34}, 722(1986).

\bibitem{17sm}
C.\ Smorra \etal,
Nature {\bf 550}, 371 (2017).

\bibitem{16dk}		
Y.\ Ding and V.A.\ Kosteleck\'y,
Phys.\ Rev.\ D {\bf 94}, 056008 (2016).

\bibitem{akgrav}
V.A.\ Kosteleck\'y,
Phys.\ Rev.\ D {\bf 69}, 105009 (2004).

\bibitem{de99}
H.\ Dehmelt, R.\ Mittleman, R.S.\ Van Dyck, Jr., and P.\ Schwinberg,
Phys.\ Rev.\ Lett.\ {\bf 83}, 4694 (1999).

\bibitem{17sc}
G.\ Schneider \etal,
Science {\bf 358}, 1081 (2017).

\bibitem{ga99}
G.\ Gabrielse, A.\ Khabbaz, D.S.\ Hall, 
C.\ Heimann, H.\ Kalinowsky, and W.\ Jhe,
Phys.\ Rev.\ Lett.\ {\bf 82}, 3198 (1999).

\bibitem{bkr97}
R.\ Bluhm, V.A.\ Kosteleck\'y, and N.\ Russell,
Phys.\ Rev.\ Lett.\ {\bf 79}, 1432 (1997).

\bibitem{bkr98}
R.\ Bluhm, V.A.\ Kosteleck\'y, and N.\ Russell,
Phys.\ Rev.\ D {\bf 57}, 3932 (1998).

\bibitem{mi99}
R.K.\ Mittleman, I.I.\ Ioannou, H.G.\ Dehmelt, and N.\ Russell,
Phys.\ Rev.\ Lett.\ {\bf 83}, 2116 (1999).

\bibitem{19ding}
Y.\ Ding, 
Symmetry {\bf 11}, 1220 (2019).

\bibitem{sm19}
C.\ Smorra \etal,
Nature {\bf 575}, 310 (2019).

\bibitem{ha00}
M.\ Hayakawa, 
Phys.\ Lett.\ B {\bf 478}, 394 (2000).

\bibitem{chklo01}
S.M.\ Carroll, J.A.\ Harvey, V.A.\ Kosteleck\'y,
C.D.\ Lane, and T.\ Okamoto,
Phys.\ Rev.\ Lett.\ {\bf 87}, 141601 (2001).

\bibitem{susy}
A.C.\ Lehum,
Europhys.\ Lett.\ {\bf 112}, 51001 (2015);
M.\ Faizal and P.A.\ Ganai,
Europhys.\ Lett.\ {\bf 111}, 21001 (2015);
H.\ Belich, L.D.\ Bernald, P.\ Gaete, J.A.\ Helay\"el-Neto, and F.J.L.\ Leal,
Eur.\ Phys.\ J.\ C {\bf 75}, 291 (2015);
J.L.\ Chkareuli,
Bled Workshops Phys.\ {\bf 15}, 46 (2014);
A.C.\ Lehum, J.R.\ Nascimento, A.Yu.\ Petrov, and A.J.\ da Silva,
Phys.\ Rev.\ D {\bf 88}, 045022 (2013);
H.\ Belich, L.D.\ Bernald, P.\ Gaete, and J.A.\ Helay\"el-Neto,
Eur.\ Phys.\ J.\ C {\bf 73}, 2632 (2013);
A.B.\ Clark, 
JHEP {\bf 1401}, 134 (2014);
C.F.\ Farias, A.C.\ Lehum, J.R.\ Nascimento, and A.Yu.\ Petrov,
Phys.\ Rev.\ D {\bf 86}, 065035 (2012);
D.\ Redigolo,
Phys.\ Rev.\ D {\bf 85}, 085009 (2012);
D.\ Colladay and P.\ McDonald,
Phys.\ Rev.\ D {\bf 83}, 025021 (2011);
P.A.\ Bolokhov, S.G.\ Nibbelink, and M.\ Pospelov,
Phys.\ Rev.\ D {\bf 72}, 015013 (2005); 
M.S.\ Berger and V.A.\ Kosteleck\'y,
Phys.\ Rev.\ D {\bf 65}, 091701(R) (2002).

\bibitem{causality}
M.\ Cambiaso, R.\ Lehnert and R.\ Potting,
Phys.\ Rev.\ D {\bf 90}, 065003 (2014);
I.T.\ Drummond,
Phys.\ Rev.\ D {\bf 88}, 025009 (2013);
M.A.\ Hohensee, D.F.\ Phillips and R.L.\ Walsworth,
arXiv:1210.2683;
M.\ Schreck,
Phys.\ Rev.\ D {\bf 86}, 065038 (2012);
F.R.\ Klinkhamer and M.\ Schreck,
Nucl.\ Phys.\ B {\bf 848}, 90 (2011);
C.M.\ Reyes,
Phys.\ Rev.\ D {\bf 87}, 125028 (2013);
Phys.\ Rev.\ D {\bf 82}, 125036 (2010);
V.A.\ Kosteleck\'y and R.\ Lehnert,
Phys.\ Rev.\ D {\bf 63}, 065008 (2001).

\bibitem{finsler}
J.E.G.\ Silva, 
arXiv:1602.07345;
J.E.G.\ Silva, R.V.\ Maluf and C.A.S.\ Almeida,
arXiv:1511.00769;
J.\ Foster and R.\ Lehnert,
Phys.\ Lett.\ B {\bf 746}, 164 (2015);
N.\ Russell,
Phys.\ Rev.\ D {\bf 91}, 045008 (2015);
M.\ Schreck,
Phys.\ Rev.\ D {\bf 91}, 105001 (2015);
Phys.\ Rev.\ D {\bf 92}, 125032 (2015);
Eur.\ J.\ Phys.\ C {\bf 75}, 187 (2015);
Phys.\ Rev.\ D {\bf 93}, 105017 (2016);
Phys.\ Rev.\ D {\bf 94}, 025019 (2016);
J.E.G.\ Silva and C.A.S.\ Almeida,
Phys.\ Lett.\ B {\bf 731}, 74 (2014); 
V.A.\ Kosteleck\'y, N.\ Russell, and R.\ Tso,
Phys.\ Lett.\ B {\bf 716}, 470 (2012);
D.\ Colladay and P.\ McDonald,
Phys.\ Rev.\ D {\bf 85}, 044042 (2012);
Phys.\ Rev.\ D {\bf 92}, 085031 (2015);
V.A.\ Kosteleck\'y,
Phys.\ Lett.\ B {\bf 701}, 137 (2011).
B.\ Edwards and V.A.\ Kosteleck\'y,
Phys.\ Lett.\ B {\bf 786}, 319 (2018).

\bibitem{muon}
R.\ Bluhm, V.A.\ Kosteleck\'y, and C.D.\ Lane,
Phys.\ Rev.\ Lett.\ {\bf 84}, 1098 (2000);
M.\ Deile \etal,
hep-ex/0110044;
G.W.\ Bennett \etal,
Phys.\ Rev.\ Lett.\ {\bf 100}, 091602 (2008);
Y.V.\ Stadnik, B.M.\ Roberts, and V.V.\ Flambaum,
Phys.\ Rev.\ D {\bf 90}, 045035 (2014).

\bibitem{gkv14}
A.H.\ Gomes, V.A. Kosteleck\'y, and A.J.\ Vargas,
Phys.\ Rev.\ D {\bf 90}, 076009 (2014).

\bibitem{kv15}
V.A.\ Kosteleck\'y and A.J.\ Vargas,
Phys.\ Rev.\ D {\bf 92}, 056002 (2015).

\bibitem{kv18}
V.A.\ Kosteleck\'y and A.J.\ Vargas,
Phys.\ Rev.\ D {\bf 98}, 036003 (2018).

\bibitem{km13}
V.A.\ Kosteleck\'y and M.\ Mewes,
Phys.\ Rev.\ D {\bf 88}, 096006 (2013).

\bibitem{19kl}	
V.A.\ Kosteleck\'y and Z.\ Li, 
Phys.\ Rev.\ D {\bf 99}, 056016 (2019).

\bibitem{km09}
V.A.\ Kosteleck\'y and M.\ Mewes,
Phys.\ Rev.\ D {\bf 80}, 015020 (2009);
Ap.\ J.\ {\bf 689}, L1 (2008).

\bibitem{km12}
V.A.\ Kosteleck\'y and M.\ Mewes,
Phys.\ Rev.\ D {\bf 85}, 096005 (2012);
J.S.\ D\'\i az, V.A.\ Kosteleck\'y, and M.\ Mewes,
Phys.\ Rev.\ D {\bf 89}, 043005 (2014).

\bibitem{nonmingrav}
V.A.\ Kosteleck\'y and M.\ Mewes,
Phys.\ Lett.\ B {\bf 757}, 510 (2016); 
Q.G.\ Bailey, V.A.\ Kosteleck\'y, and R.\ Xu,
Phys.\ Rev.\ D {\bf 91}, 022006 (2015);
V.A.\ Kosteleck\'y and J.D.\ Tasson,
Phys.\ Lett.\ B {\bf 749}, 551 (2015).

\bibitem{km12}
V.A.\ Kosteleck\'y and M.\ Mewes,
Phys.\ Rev.\ D {\bf 85}, 096005 (2012).

\bibitem{anti}
The antisymmetry of the index pair ``12" on the right sides 
of the definitions \rf{tilds} and \rf{jjs}
is reflected by the spin index pair $\mn$ of the relevant coefficients 
together with the form of the expression.
For example,
the definition of $\bptw w {3}$ in \rf{tilds}
contains a term of 
$g_{w}^{120} - g_{w}^{012} + g_{w}^{021}$.
$g_{w}^{120}$ is antisymmetric on the index pair ``12"
from the property of the  $g_{w}^{\mn\al}$ coefficient
as the first two indices are contracted with the antisymmetric
Dirac matrix $\si_\mn$.
The remaining expression $- g_{w}^{012} + g_{w}^{021}$ 
is antisymmetric on the index pair ``12" by observation. 
Similar arguments can be made for other terms containing 
the index pair ``12" to prove that it is antisymmetric in these definitions. 

\bibitem{sunframe0}
V.A.\ Kosteleck\'y and C.D.\ Lane,
Phys.\ Rev.\ D {\bf 60}, 116010 (1999).

\bibitem{sunframe}
V.A.\ Kosteleck\'y and M.\ Mewes,
Phys.\ Rev.\ D {\bf 66}, 056005 (2002).

\bibitem{ycon}
The common adopted ``$y$-convention'' for the Euler angles $(\al,\be,\ga)$ 
from coordinate system $(x,y,z)$ to $(x^1,x^2,x^3)$  is defined as follows. 
First, rotate $xyz$ counterclockwise around its $\hat{z}$ axis by $\al$
to give $x^\prime y^\prime z^\prime$. 
Then, rotate $x^\prime y^\prime z^\prime$ counterclockwise around its $\hat{y}’$ axis by $\be$ 
to give $x^{\prime\prime}y^{\prime\prime}z^{\prime\prime}$. 
Finally, rotate $x^{\prime\prime}y^{\prime\prime}z^{\prime\prime}$ 
counterclockwise around its $\hat{z}^{\prime\prime}$ axis by $\ga$ to give $x^1x^2x^3$. 

\bibitem{ca04}
F.\ Can\`e, D.\ Bear, D.F.\ Phillips, M.S.\ Rosen, C.L.\ Smallwood, R.E.\ Stoner, R.L.\ Walsworth, and V.A.\ Kosteleck\'y,
Phys.\ Rev.\ Lett.\ {\bf 93}, 230801 (2004).

\bibitem{he08}
B.R.\ Heckel, E.G.\ Adelberger, C.E.\ Cramer, T.S.\ Cook, S.\ Schlamminger, and U.\ Schmidt,
Phys.\ Rev.\ D {\bf 78}, 092006 (2008).

\bibitem{space}
R.\ Bluhm, V.A.\ Kosteleck\'y, C.D.\ Lane, and N.\ Russell, 
Phys.\ Rev.\ D {\bf 68}, 125008 (2003);
Phys.\ Rev.\ Lett.\ {\bf 88}, 090801 (2002).



\end{thebibliography}
\end{document}